\documentclass[10pt,journal,compsoc]{IEEEtran}

\usepackage{multirow}
\usepackage{subfigure}
\usepackage[linesnumbered,ruled,vlined]{algorithm2e}
\usepackage{booktabs}
\usepackage{amsmath}
\usepackage{color}

\usepackage{amssymb}
\usepackage{amsfonts}
\usepackage{epsfig}
\usepackage{graphicx}
\usepackage{epstopdf}
\usepackage{enumitem}
\usepackage{times}
\usepackage{xcolor}
\usepackage{soul}
\usepackage{float}
\usepackage{mathrsfs}
\usepackage{amssymb}
\usepackage{psfrag}
\usepackage{stfloats}
\usepackage{float}
\usepackage{comment}
\usepackage{array}
\usepackage{url}
\usepackage{caption}
\usepackage{amsthm}
\usepackage{booktabs}  %
\ifCLASSOPTIONcompsoc
  \usepackage[nocompress]{cite}
\else
  \usepackage{cite}
\fi
\ifCLASSINFOpdf
\else
\fi
\begin{document}
%
\title{Attention over Self-attention:\\ Intention-aware Re-ranking with Dynamic Transformer Encoders for Recommendation}
%
%
%
%

\author{Zhuoyi~Lin, 
        Sheng~Zang,
        Rundong~Wang,
        Zhu~Sun,
        J. Senthilnath, 
        Chi~Xu,
        and~Chee Keong~Kwoh
\IEEEcompsocitemizethanks{\IEEEcompsocthanksitem Zhuoyi Lin is with the Institute for Infocomm Research, Agency for Science, Technology and Research (A*STAR), Singapore, and the School of Computer Science and Engineering, Nanyang Technological University, Singapore. (Email: zhuoyi001@e.ntu.edu.sg)

\IEEEcompsocthanksitem Sheng Zang is with the college of computer science, Nankai University, China. (Email: 1813048@mail.nankai.edu.cn).

\IEEEcompsocthanksitem  Rundong Wang, and Chee Keong Kwoh are with the School of Computer Science and Engineering, Nanyang Technological University, Singapore (E-mail:	rundong001@e.ntu.edu.sg,
asckkwoh@ntu.edu.sg).

\IEEEcompsocthanksitem Zhu Sun (corresponding author) is with Institute of High Performance Computing and Centre for
Frontier AI Research, A*STAR, Singapore. (E-mail:sunzhuntu@gmail.com).

\IEEEcompsocthanksitem J. Senthilnath is with the Institute for Infocomm Research, Agency for Science, Technology and Research (A*STAR), Singapore. (E-mail: J\_Senthilnath@i2r.a$-$star.edu.sg).

\IEEEcompsocthanksitem Chi Xu is with the Singapore Institute of Manufacturing Technology, A*STAR, Singapore, and the School of Computer Science and Engineering, Nanyang Technological University, Singapore. (Email: cxu@simtech.a-star.edu.sg). 
}
}

\IEEEtitleabstractindextext{%
\begin{abstract}
Re-ranking models refine item recommendation lists generated by the prior global ranking model, which have demonstrated their effectiveness in improving the recommendation quality.
However, most existing re-ranking solutions only learn from implicit feedback with a shared prediction model, which regrettably ignore inter-item relationships under diverse user intentions.
In this paper, we propose a novel Intention-awa\underline{r}e Re-r\underline{a}nking Model with Dynam\underline{i}c Tran\underline{s}former \underline{E}ncoder {(RAISE)}, aiming to perform user-specific prediction for each individual user based on her intentions. Specifically, we first propose to mine latent user intentions from text reviews with an intention discovering module (IDM). 
By differentiating the importance of review information with a co-attention network, the latent user intention can be explicitly modeled for each user-item pair.
We then introduce a dynamic transformer encoder (DTE) to capture user-specific {inter-item} relationships among item candidates by seamlessly accommodating the learned latent user intentions via IDM.
As such, one can not only achieve more personalized recommendations but also obtain corresponding explanations by constructing RAISE upon existing recommendation engines.
Empirical study on four public datasets shows the superiority of our proposed RAISE, with up to 13.95\%, 9.60\%, and 13.03\% relative improvements evaluated by Precision@5, MAP@5, and NDCG@5 respectively.
\end{abstract}

}

\maketitle

\IEEEdisplaynontitleabstractindextext

%
\IEEEpeerreviewmaketitle



%
%
%
%

\section{Introduction}
\IEEEPARstart{I}{n} the era of big data, recommender systems are widely adopted by the online platforms (e.g., Amazon and Youtube), so as to alleviate the problem of information overload~\cite{sun2019research,lin2021glimg}.
Accordingly, latent factor models, e.g., matrix factorization~\cite{koren2009matrix,lin2020comet}, and deep learning models, e.g., NeuMF~\cite{NCF}, have demonstrated their effectiveness to achieve personalized recommendations by learning user and item representations.
Despite the great success, one fundamental assumption of the above solutions is that a global ranking model is designed to optimize the overall performance of item recommendations. This could be sub-optimal for individual users because it ignores the local item distributions for each user~\cite{DLCM,PRM}. 
\begin{figure}[t]
\centering
\includegraphics[width=0.45\textwidth]{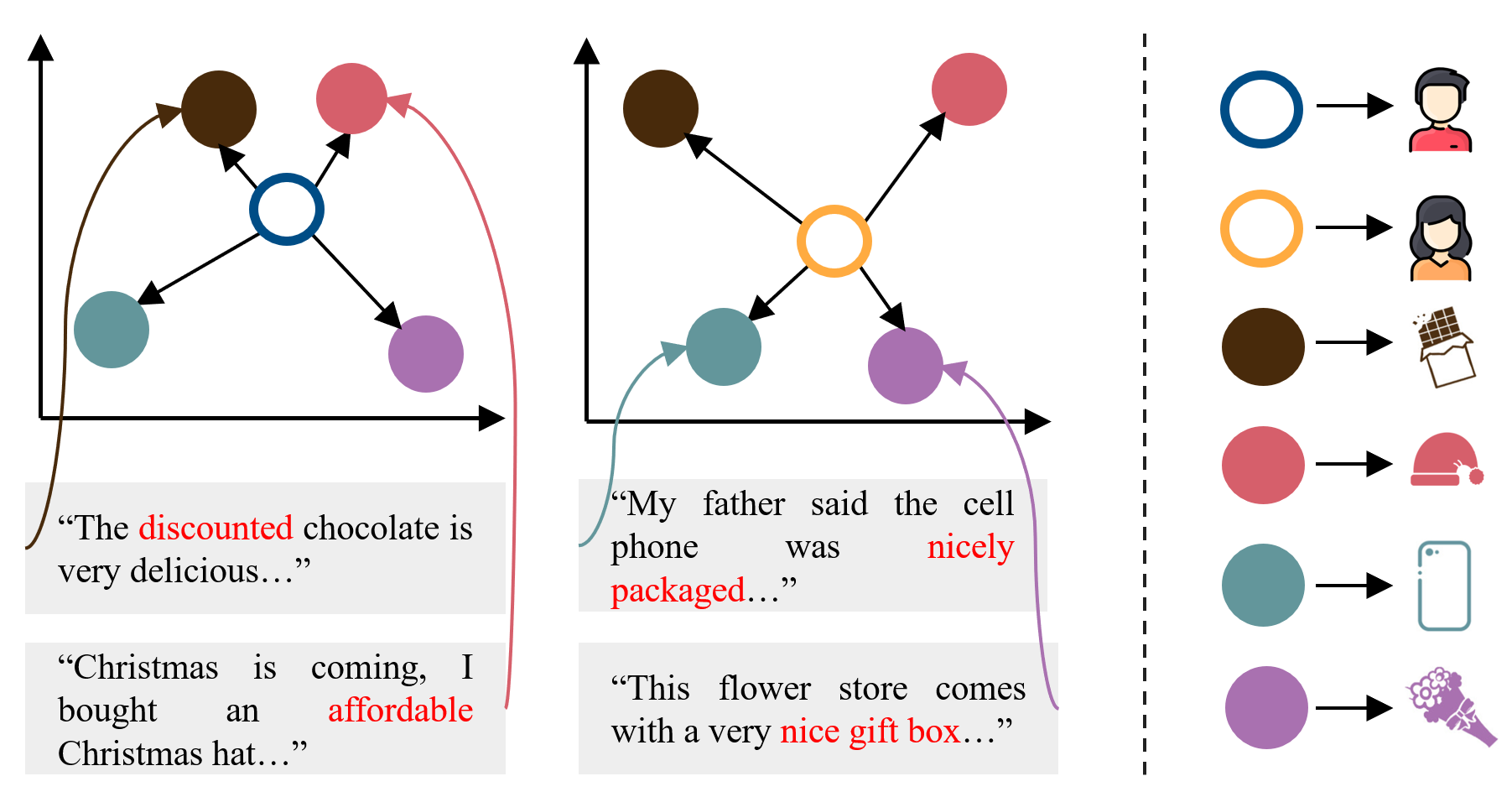} %
\caption{An illustration of users' complex intentions in the latent space. The left user can be price-sensitive and hence cheap items (i.e., chocolate and Christmas hat) are closer in the latent space, while the right user may prefer the nicely packaged items so that the representations of cell phone and bouquet are more similar.
}\label{fig:user_intention}
\vspace{-0.1in}
\end{figure}
To remedy the above issue, a shared re-ranking model is employed to refine the initial recommendation lists provided by the prior global ranking model for each user. 
Specifically, by taking into account the relationships among item candidates in a user's initial recommendation list, the {inter-item} relationships for the target user can be explicitly modeled \cite{DLCM,PRM,liu2020personalized}. 
As such, re-ranking models are able to re-rank the item candidates and generate more personalized recommendation list for each target user. However, we argue that the existing manner of modeling {inter-item} patterns is insufficient, as such item relationships could vary a lot under different user intentions\footnote{Following Wang et al. \cite{wang2020disentangled}, user intention is a high-level concept, which could be the reasons of a user to purchase a certain item (e.g., timing, price, user interest and shopping for others). Note that we focus on refining the item recommendation lists with learnt latent user intentions from text reviews rather than predicting the next item with instant intention for the target user (e.g., {sequential} recommendation models such as BERT4Rec~\cite{bert4rec} and IMfOU~\cite{IMfOU}).}.
Intuitively, users have multiple intentions to adopt certain items; meanwhile, different user intentions could drive different user behaviors and item relationships \cite{wang2020disentangled, chen2019air}. Taking Fig.~\ref{fig:user_intention} as an example, the representations of two cheap items could be similar for a price-sensitive user, while the representations of the two aforementioned items could be different for a user who often buys gifts for her friends, as she may be more concerned with whether items are well packaged. Besides, although the {existing} re-ranking models can capture the {inter-item} relationships among item candidates, they are shared by all users in a dataset, which may not be sufficiently `personalized'.

Consequently, we are seeking to investigate the re-ranking task through modeling user-specific {inter-item} relationships based on user intentions. However, this is not trivial because of two main challenges: 
(1) Basically, user intentions are diverse and complex, which may vary greatly when confronted with different items.
How to accurately capture user intentions is of crucial importance to deliver a performance-enhanced re-ranking model;
(2) To provide sufficient personalization, a shared re-ranking model is not feasible. However, it is impractical to assign each user a prediction model. Hence, a {tailored} solution need to be designed in order to achieve both {effectiveness and efficiency}. 

To tackle these challenges, we propose a novel Intention-awa\underline{r}e Re-r\underline{a}nking Model with Dynam\underline{i}c Tran\underline{s}former \underline{E}ncoder
{(RAISE)}. In particular, we first devise an \emph{\textbf{intention discovering module}} (IDM) to mine latent user intentions from text reviews. As illustrated in Fig.~\ref{fig:user_intention}, such auxiliary information contains users' preferences and item properties, which could be useful for modeling user intentions and item relationships \cite{mcauley2013hidden,bao2014topicmf}. 
Given a user-item pair, IDM applies a co-attention network to estimate review-to-review matching scores and differentiate users' diverse intentions from text reviews.
This enables intention-aware representations to be generated by weighting the text representations with learnt matching scores.
We then design a \emph{\textbf{dynamic transformer encoder}} (DTE) to perform user-specific predictions by seamlessly accommodating the learnt latent user intentions.
Under the hood, the dynamic self-attention mechanism captures the user-specific {inter-item} relationships and provides the driving force: an individual attention network is applied over the self-attention layer, to contextualize the item representations based on the learnt user intentions.
Our proposed DTE advances the classic transformer encoder by learning specialized transformations of input item sequences, which increases the representational capability with limited extra computational cost and keeps efficient inference.
To summarize, this paper makes the following contributions.
\begin{itemize}[leftmargin=*]
    \item We emphasize the importance of modeling diverse user intentions for the re-ranking task, whereby an IDM is devised to help extract user intentions from text reviews.
    \item We design a DTE to explicitly capture the user-specific {inter-item} patterns based on the learnt user intentions via IDM. By applying an individual attention network over the self-attention layer, DTE enables our proposed RAISE to perform user-specific predictions in an efficient manner.
    \item We conduct extensive experiments and ablation studies on four public datasets to verify the effectiveness and interpretability of our proposed RAISE.
\end{itemize}
\section{Related Works}
This section first briefly reviews existing re-ranking studies.
Since our proposed RAISE aims to perform user-specific prediction based on the input item sequences and user intentions, we then present existing review-aware recommendation methods and input-dependent recommendation models.

\begin{figure*}[t]
\centering
      \subfigure[Scaled dot-product attention.]{
    \begin{minipage}{0.2\textwidth}
      \centering
      \includegraphics[width=1.1in]{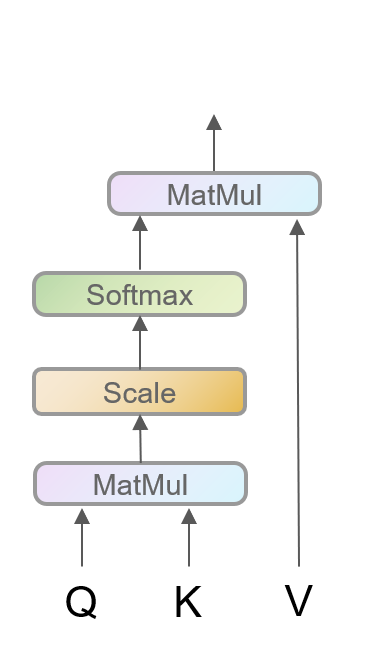}
    \end{minipage}}%
      \subfigure[Multi-head self-attention.]{
    \begin{minipage}{0.3\textwidth}
      \centering
      \includegraphics[width=1.7in]{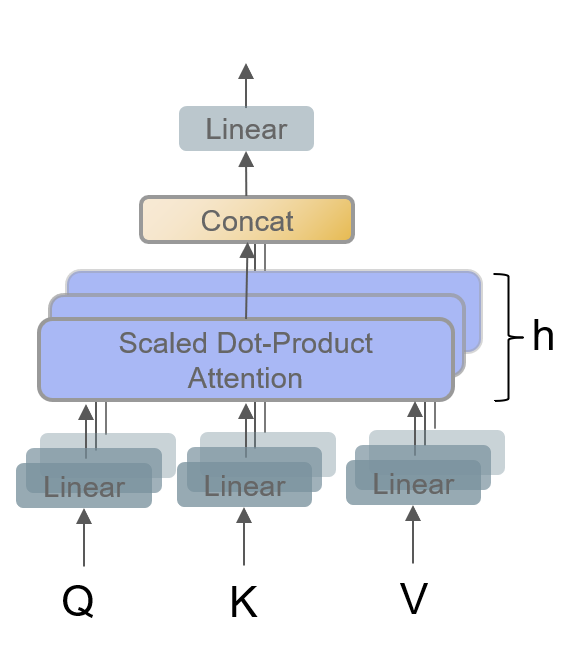}
    \end{minipage}}
      \subfigure[Our dynamic self-attention.]{
    \begin{minipage}{0.45\textwidth}
      \centering
      \includegraphics[width=3.in]{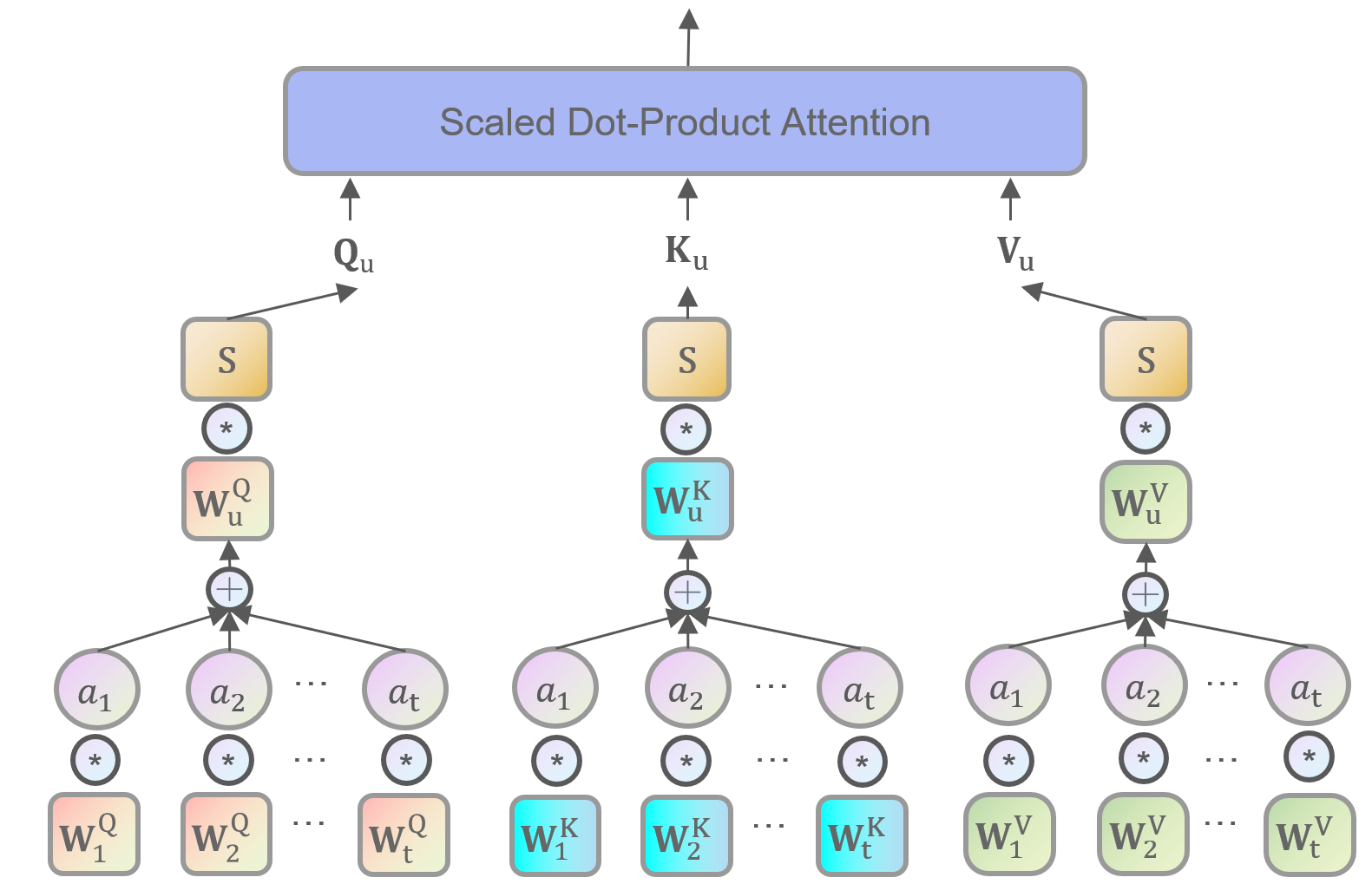} 
      \vspace{0.07in}
    \end{minipage}}
    \vspace{-0.1in}
  \caption{ Illustration of dynamic self-attention, which is the key design of dynamic transformer encoder.}
    \label{fig:dynamic_self_attention}
\end{figure*}

%
\smallskip\noindent\textbf{Re-ranking Models for Recommendation.}
{Re-ranking methods have demonstrated their effectiveness to improve user satisfaction and been widely applied in industrial applications.
For example, diversity-aware re-ranking methods focus on presenting relevant but diverse results at the top of a ranked list \cite{yan2021diversification, abdool2020managing}, while fairness-aware re-ranking methods aim to provide each items a fair proportion of exposure \cite{oosterhuis2021computationally, zhu2021fairness, yadav2021policy}.} 
{In this work, we aim to capture the mutual relationships between items in the initial recommendation list for each user in order to improve model accuracy.}
For example, DLCM~\cite{DLCM} uses gated recurrent unit (GRU) to sequentially encode the information of top candidates into document representations, and SetRank~\cite{pang2020setrank} {uses multi-head self-attention mechanisms and an attention rank loss function to re-rank documents for the document retrieval task.}
In addition, Seq2Slate~\cite{bello2018seq2slate} exploits mutual influences between items with pointer networks, which treats the recommendation task as a sequence generation problem.
Meanwhile, a graph neural network (GNN) based re-ranking method IRGPR~\cite{liu2020personalized} has been proposed to fuse the information from multi-hop neighbors and item relationships.
It essentially models the global user-item and item-item relationships rather than the local item-item relationships among the initial recommendation lists.
{A more recent work \cite{yan2022personalized} propose to combine graph attention networks (GAT) and transformer model to capture the sequential signals underlying users' behavior sequences for complementary product recommendation. Then it uses a hinge loss to perform re-ranking for sequential recommendation tasks without a specific model architecture.}
The most related work to ours is PRM~\cite{PRM}. 
By employing transformer encoders to encode the input items, the mutual influences between item candidates can be captured to refine the initial recommendations. However, the transformer encoders are shared by all users in PRM, which ignores user intentions and could lead to sub-optimal performance.

To sum up, existing works follow the same paradigm of modeling the {inter-item} relationships to re-rank the initial recommendation and regrettably ignore diverse user intentions. 
By contrast, we focus on exploiting latent user intentions from text reviews to capture the {inter-item} relationships specifically for each user in an effective manner. 
This enables RAISE to provide user-specific prediction and achieve more personalized recommendation.

\smallskip\noindent\textbf{Review-aware Recommendation Methods.}
Our work benefits from review-aware recommendation models which aim to exploit richer semantic information from the text reviews. 
Early studies such as HFT and TopicMF \cite{mcauley2013hidden,bao2014topicmf} demonstrate that better rating prediction accuracy can be achieved by modeling text reviews with topic models. 
Empowered by the powerful representational capabilities of deep neural networks, deep learning-based recommendation models such as DeepCoNN, TransNets, and MPCN \cite{zheng2017joint, catherine2017transnets, tay2018multi} are proposed to predict missing ratings from text reviews with convolutional neural networks (CNN) \cite{kim-2014-convolutional} and attention networks \cite{vaswani2017attention}.
As for ranking tasks, TAFA \cite{zhou2020tafa} provides recommendations by jointly learning representations from user reviews and implicit feedback.
Recently, TIM \cite{pena2020combining} models users and items within the topic space which is learned from the review data.
Another direction of this field is to provide explanations for item recommendations based on text reviews \cite{zhang2020explainable,sun2020DualLearning,le2021synthesizing}.
Different from the above works which mainly focus on to tackle rating prediction tasks or ranking tasks with text reviews, we focus on distinguishing user intentions from reviews so as to construct user-specific re-ranking model.

\smallskip\noindent\textbf{Input-dependent Models.} Recently, input-dependent models have shown effectiveness 
in various domains, such as language modeling~\cite{wu2019pay, peng2020mixture} and computer vision~\cite{yang2019condconv,chen2020dynamic}.
In recommendation, IFM~\cite{yu2019input} and DIFM~\cite{lu2020dual} are presented to re-weight the representations of features and weights for different input instances before performing feature interactions. Inspired by these studies, we design a dynamic transformer encoder which performs an individual attention network on the self-attention layer and enables modeling user-specific {inter-item} patterns unveiled by the user intentions. We demonstrate that the proposed dynamic transformer encoder is computionally efficient with superior performance in this study.   
\section{Preliminaries}
\textbf{Notations and Problem Formulation.}
Given a set of users $\mathcal{U}=\{u_1,u_2,\cdots\}$, a set of items $\mathcal{I}=\{i_1,i_2,\cdots\}$, and observed interaction scores $y_{ui}$. we adopt GMF~\cite{NCF} as the prior global ranking model, which predicts the missing interaction score ($\tilde{y}_{ui}$) between the target user $u$ and target item $i$ with their learnt representations denoted as $\mathbf{p}_u$ and $\mathbf{q}_i$, respectively.
In this way, an initial recommendation list $\mathcal{S}_{u}=[i_{1},i_{2},\cdots, i_{n}]$ can be generated by ranking the estimated interaction scores for user $u$, where $n$ is the length of $\mathcal{S}_{u}$.
Given $\mathcal{S}_{u}$, $\mathbf{p}_u$ and $\{\mathbf{q}_i|i\in\mathcal{S}_u\}$ learnt from GMF, 
the goal of our study is essentially how to {effectively} capture the {inter-item} relationships under different user intentions from text reviews and {efficiently} perform user-specific refinement for $\mathcal{S}_{u}$. 
Hence, we further introduce two review sequences $\mathcal{R}_{u}=\{r_{1}^{(u)}, r_{2}^{(u)}, \cdots, r_{\textit{l}_u}^{(u)} \}$ and  $\mathcal{R}_{i}= \{r_{1}^{(i)}, r_{2}^{(i)}, \cdots, r_{\textit{l}_i}^{(i)} \}$ for user $u$ and item $i$, which contains the reviews written by user $u$ and reviews received by item $i$, respectively. Note that $\textit{l}_u$ and $\textit{l}_i$ are the maximum number of reviews of $u$ and $i$. 
As such, each user-item pair in the training set can be denoted as a {6-tuple} ($u,i,\mathbf{p}_{u},\mathbf{q}_{i},\mathcal{R}_{u}, \mathcal{R}_{i}$), and RAISE is trained to re-rank the item candidates in $\mathcal{S}_{u}$ for user $u$.

\begin{table}[!t]
\scriptsize
\centering
\caption{Main mathematical notations used in RAISE.}
\vspace{-0.1in}
\begin{tabular}{|@{}c|c@{}|}
\toprule
Notations                      & Definitions and Descriptions                                                                                                     \\ \midrule
$\mathbf{p}_u$, $\mathbf{q}_i$   & \begin{tabular}[c]{@{}c@{}}Latent representations of user $u$ and item $i$ which \\ are learnt by GMF from implicit feedback.\end{tabular} \\
$\mathcal{S}_u$                         & \begin{tabular}[c]{@{}c@{}}Initial item recommendation list learnt by GMF \\ for user $u$. \end{tabular}                             \\
$\hat{y}_{ui}$ & Prediction score for user $u$ to item $i$.                                                                                               \\

$\mathcal{R}_u, \mathcal{R}_i$                   & \begin{tabular}[c]{@{}c@{}}User $u$ and item $i$'s review sequences.\end{tabular}                  \\
$c_{kj}$                        & \begin{tabular}[c]{@{}c@{}}Matching score between user's $k$th review \\ and item's $j$th review.\end{tabular}                        \\
$\mathbf{r}_{i}^{(u)}$, $\mathbf{r}_{i}^{(i)}$      & \begin{tabular}[c]{@{}c@{}}Intention-aware review representations \\ for $u$-$i$ pair.\end{tabular}                          \\
$\mathbf{o}_i$                        & Position embedding of item $i$.                                                                                                     \\
$\mathbf{s}_{i}$ &   \begin{tabular}[c]{@{}c@{}} Item $i$'s final representation which is \\ the input of the dynamic transformer encoders. \end{tabular}                                                                             \\
$\mathbf{Q}_u, \mathbf{K}_u, \mathbf{V}_u$             & User-specific query, key, and value representations.                                                                              \\
$\mathbf{W}_u^Q, \mathbf{W}_u^K, \mathbf{W}_u^V$     & User-specific transform matrices.                                                                                                 \\
$\mathbf{a}_{t}$                            &               
\begin{tabular}[c]{@{}c@{}} Shared attention weights in the dynamic \\ transformer encoder.  \end{tabular}       
\\ \bottomrule
\end{tabular}
\vspace{-0.15in}
\end{table}
\begin{figure*}[t]
\centering
      \subfigure[Architecture of RAISE.]{
    \begin{minipage}{0.5\textwidth}
      \centering
      \includegraphics[width=3.5in]{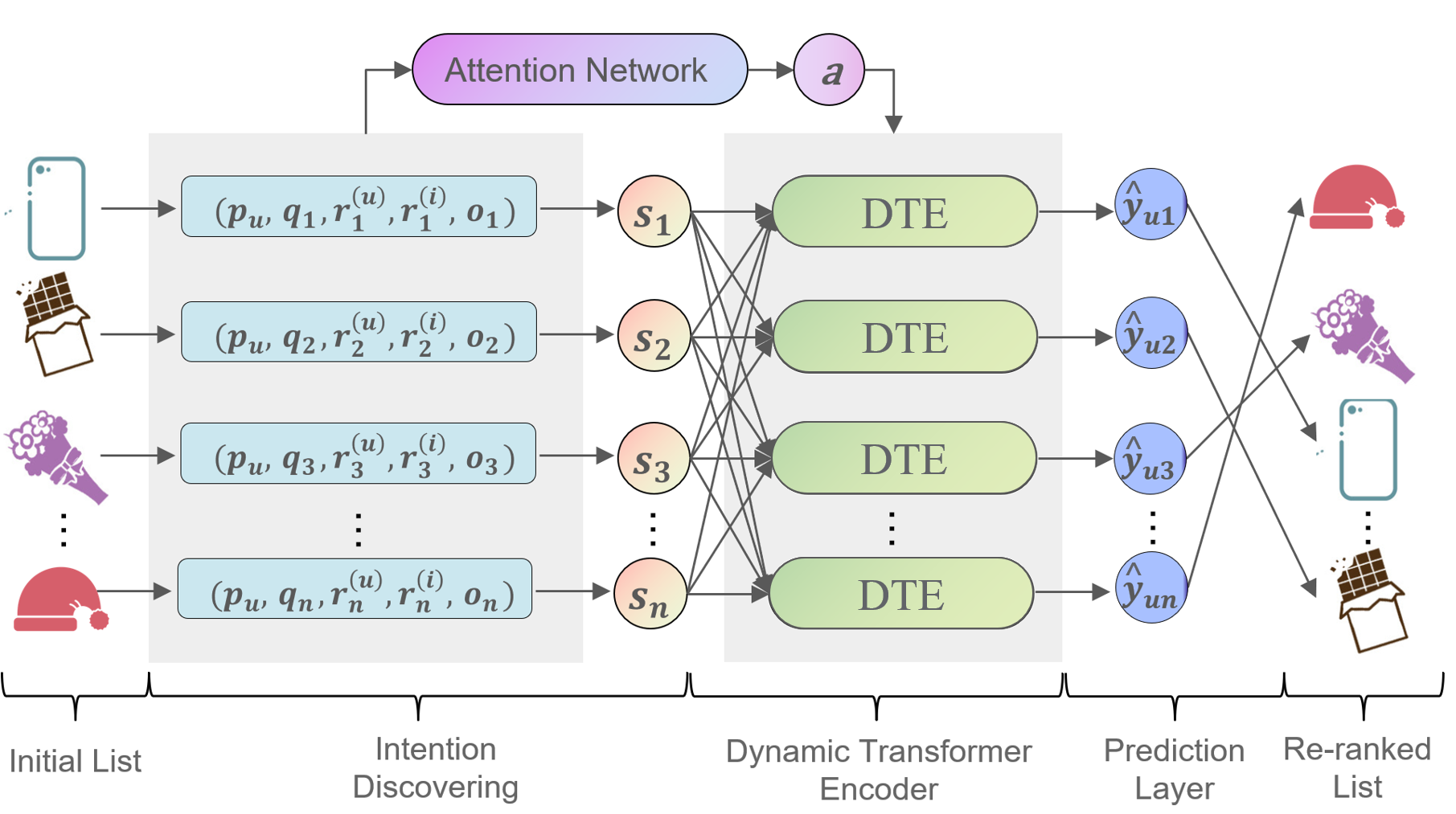}
    \end{minipage}}%
      \subfigure[DTE.]{
    \begin{minipage}{0.2\textwidth}
      \centering
      \includegraphics[width=1.08in]{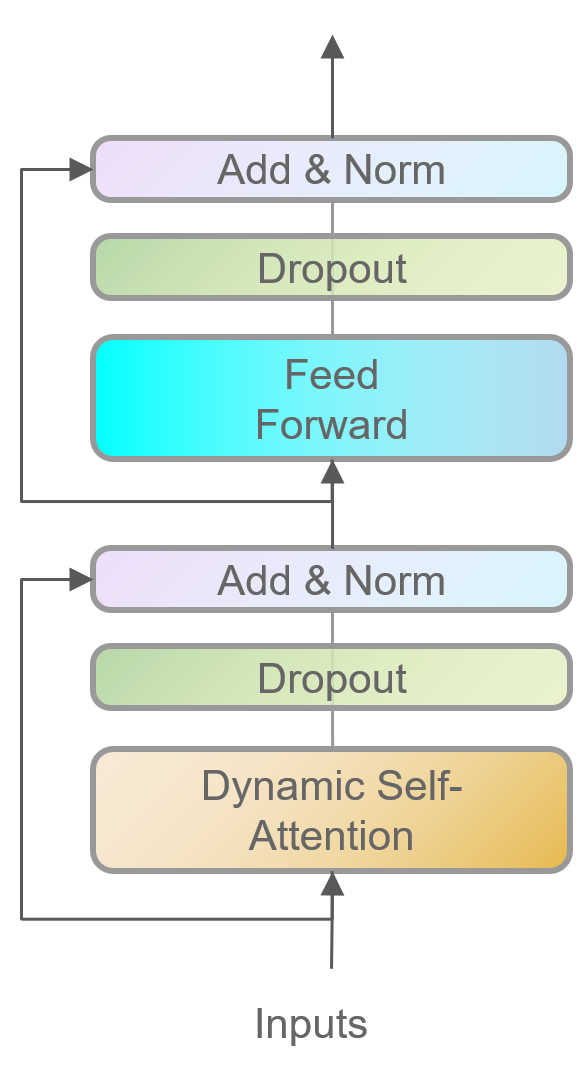}
    \end{minipage}}
      \subfigure[Co-attention Module.]{
    \begin{minipage}{0.26\textwidth}
      \centering
      \includegraphics[width=2.05in]{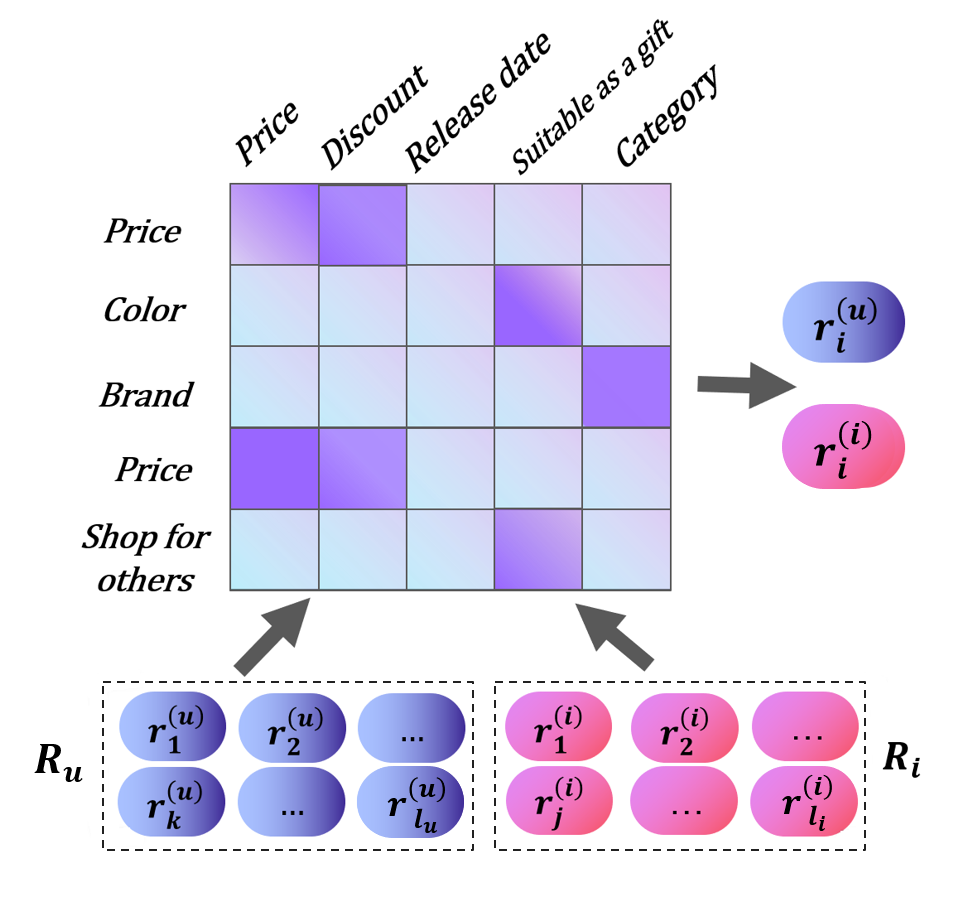}
    \end{minipage}}
    \vspace{-0.1in}
  \caption{ Overall framework of RAISE.}
  \label{fig:framework} 
  \vspace{-0.15in}
\end{figure*}

\smallskip\noindent\textbf{Self-attention Mechanism.}
Our work benefits from the self-attention mechanism due to its effectiveness to capture the {inter-item} patterns among item candidates \cite{vaswani2017attention}. 
As shown in Fig. \ref{fig:dynamic_self_attention}(a). After linearly projecting the input sequence $\mathbf{S}$ to obtain queries ($\mathbf{Q}$) and keys ($\mathbf{K}$) of dimension $d_k$, and values ($\mathbf{V}$) of dimension $d_v$, the attention function produces outputs of dimension $d$, which is defined as follows:
\begin{equation}\label{equ:self_attention}
\footnotesize
\begin{aligned}
& Attention(\mathbf{Q},\mathbf{K},\mathbf{V}) = Softmax(\frac{\mathbf{Q}\mathbf{K}^{T}}{\sqrt{d_{k}}})\mathbf{V} \\
& \text{where }  \mathbf{Q} = \mathbf{S}\mathbf{W}^{Q} \text{,  } \mathbf{K} = \mathbf{S}\mathbf{W}^{K} \text{, and } \mathbf{V} = \mathbf{S}\mathbf{W}^{V}
\end{aligned}
\end{equation}
where $\mathbf{W}^{Q}$, $\mathbf{W}^{K}$, and $\mathbf{W}^{V}$ are transform matrices. 

In addition, the self-attention mechanism can be multi-headed by employing a number of heads $h$ ~\cite{vaswani2017attention,peng2020mixture}. As shown in Fig. \ref{fig:dynamic_self_attention}(b), each head learns separate attention weights from different representation subspaces:
\begin{equation} \label{multi-head selfattention}
\footnotesize
\begin{aligned}
& \text{Multi-head}(\mathbf{Q}, \mathbf{K}, \mathbf{V}) = \text{Concat}(head_{1}, \cdots, head_{h})\mathbf{W}^{O}\\
& \text{where }  head_{i} = \text{Attention}(\mathbf{Q}\mathbf{W}_{i}^{Q}, \mathbf{K}\mathbf{W}_{i}^{K}, \mathbf{V}\mathbf{W}_{i}^{V})
\end{aligned}
\end{equation}
where $\mathbf{W}^{O}\in\mathbb{R}^{hd_{v} \times d}, \mathbf{W}^{Q}_{i}\in\mathbb{R}^{d \times d_{k}},  \mathbf{W}^{K}_{i}\in\mathbb{R}^{d \times d_{k}}, \mathbf{W}^{V}_{i}\in\mathbb{R}^{d \times d_{v}}$ are transform matrices used in the multi-head self-attention.
Normally, $d_k$ and $d_v$ are set to $d/h$. 
The extra computational cost of multi-head self-attention is limited compared with that of normal self-attention function with full dimensionality (see Eq.~\ref{equ:self_attention}).
We will provide a more detailed discussion in {Dynamic Self-attention Analysis}.

\section{The Proposed Method}
In this section, we first briefly introduce the architecture of our proposed re-ranking method {RAISE}, then we present each component of RAISE in detail. 

\subsection{Overall of RAISE}
The overall architecture of RAISE is presented in Fig.~\ref{fig:framework}(a), which consists of \emph{intention discovering module} (IDM), \emph{dynamic transformer encoders} (DTE), and \emph{prediction layer}. Given an initial list $\mathcal{S}_{u}$ and learnt representations $\mathbf{p}$ and $\mathbf{q}$ from GMF, the goal of RAISE is to refine $\mathcal{S}_{u}$  by modeling the local {inter-item} relationships based on user intentions. To this end, IDM first discovers user intentions from text review with a co-attention module. The learnt intention-aware representations from text review are then fed into the DTE to capture user-specific item relationships. Accordingly, a new predicted score for each item candidate in $\mathcal{S}_{u}$ is generated via the prediction layer. Finally, a re-ranked item list can be achieved by ranking the predicted scores.
\subsection{Intention Discovering Module (IDM)}
\subsubsection{\textbf{Intention-aware Review Representation}}
Learning informative user intentions is of crucial importance for RAISE to perform more personalized re-ranking. Intuitively, text reviews written by users are usually semantic and encode users' diverse intentions, which could lead to better modeling of user behavior when confronted with an item \cite{tay2018multi,zheng2017joint}.

We hence employ a co-attention module to generate two intention-aware representations for each user-item pair with regard to the diverse intentions behind their reviews, as illustrated in Fig.~\ref{fig:framework}(c).  
Specifically, given two input review sequences $\mathcal{R}_{u}$ and $\mathcal{R}_{i}$, we first obtain the representation of each single text review by simply adding its constituent word representations. 
As such, two embedding sequences $\mathcal{R}_{u}^e\!\!=\!\!\{\mathbf{r}_{1}^{(u)}, \mathbf{r}_{2}^{(u)}, \cdots, \mathbf{r}_{\textit{l}_u}^{(u)} \}$ and $\mathcal{R}_{i}^e\!\!=\!\! \{\mathbf{r}_{1}^{(i)}, \mathbf{r}_{2}^{(i)}, \cdots, \mathbf{r}_{\textit{l}_i}^{(i)} \}$ can be achieved.
To distinguish the importance of reviews according to the intentions behind them, we then compute the review-level matching scores between every user-item review pair with a co-attention network, given by, 
\begin{equation}\label{equ:co-attention}
\footnotesize
 c_{kj} = f(\mathbf{r}_{k}^{(u)})^{T} \mathbf{M} f(\mathbf{r}_{j}^{(i)})
\end{equation}
where the matching matrix $\mathbf{C}=[c_{kj}]$ indicates how well the intentions behind a user review and an item review matches each other; $\mathbf{M}\in\mathbb{R}^{d\times d}$ is a trainable transform matrix; $\mathbf{r}_{k}^{(u)}, \mathbf{r}_{j}^{(i)} \in\mathbb{R}^d$ denote the representations of the $k$-th review of user $u$ and the $j$-th review of item $i$, respectively. 
Meanwhile, we employ two MLPs to encode the review representation denoted as $f(\cdot)$ in Eq.~\ref{equ:co-attention}.
Since each row (column) of $\mathbf{C}$ indicates how much the main intention of a user (item) review matches those of all item (user) reviews, the refined representations of a text review regarding users and items can be weighted by the strength of its main intention,
\begin{equation}
\footnotesize
\begin{aligned}
 &\mathbf{r}_{k}^{(u)'} = \frac{1}{|\mathcal{R}_{i}^e|}  \sum\nolimits_{j} c_{kj} * \mathbf{r}_{k}^{(u)} ; \\
 & \mathbf{r}_{j}^{(i)'} = \frac{1}{|\mathcal{R}_{u}^e|}  \sum\nolimits_{k} c_{kj} * \mathbf{r}_{j}^{(i)}.
\end{aligned}
\end{equation}
Then we add up all the reviews of user $u$ and item $i$ respectively in order to obtain the intention-aware representations $\mathbf{r}_{i}^{(u)}$ and $\mathbf{r}_{i}^{(i)}$. Despite the simplicity, it is an {effective} way to encode the quantity of each intention behind a review sequence into the final intention-aware representations,
\begin{equation}\label{equ:intention-representation}
\footnotesize
\mathbf{r}_{i}^{(u)} = \sum\nolimits_{k} \mathbf{r}_{k}^{(u)'}; \;\; \mathbf{r}_{i}^{(i)} = \sum\nolimits_{j}  \mathbf{r}_{j}^{(i)'}.
\end{equation}

Note that for each user-item pair, their {intention-aware representations} are unique and will be different w.r.t. other users or items. In other words, the intention-aware representations are contextually learnt according to the input review sequences.
This enables RAISE to capture {inter-item} patterns based on the matched user intentions between the review sequences of the target user $u$ and those of item candidates in $\mathcal{S}_{u}$, leading to more personalized recommendations.

\subsubsection{\textbf{Intention-aware Sequential Representation}}
Before introducing the DTE, a meaningful input sequential representation $\mathbf{S}$ should be generated in advance, to encode sequential item candidates in the initial list $\mathcal{S}_{u}$. 
We notice that most of the existing re-ranking models capture {inter-item} relationships from the implicit feedback data \cite{PRM,bello2018seq2slate,zhuang2018globally}. However, implicit feedback only indicate if a user has interacted with an item, thus it could be hard to learn effective item representations from implicit feedback alone. 
Besides, the learnt item representations from implicit feedback are shared by all users, which is unreasonable as item representations may vary greatly under different user intentions. 
Hence, we propose to represent item candidates with both the implicit feedback data and text review data. 
To this end, the user-specific item representations $\mathbf{s}_i$ can be achieved with latent representations learnt by GMF and intention-aware representations obtained in the IDM:
\begin{equation}\label{equ:item-representation}
\footnotesize
\begin{aligned}
& \mathbf{s}_{i} = \mathbf{W}^{S}[Concat(\mathbf{s}_{i}^{(im)}, \mathbf{s}_{i}^{(re)})] + \mathbf{o}_{i}\\
& \mathbf{s}_{i}^{(im)} = f(Concat(\mathbf{p}_{u}, \mathbf{q}_{i}))\\
& \mathbf{s}_{i}^{(re)} = f(Concat(\mathbf{r}_{i}^{(u)}, \mathbf{r}_{i}^{(i)}))
\end{aligned}
\end{equation}
where $\mathbf{W}^{S}$ is a projection matrix; $\mathbf{s}_{i}^{(im)}$ and $\mathbf{s}_{i}^{(re)}$ denote item representations learnt from the implicit feedback data via GMF and text review information via IDM, respectively;
inspired by PRM~\cite{PRM}, we also encode the initial positions of item candidates in $\mathcal{S}_{u}$ denoted as $\mathbf{o}_i$. 
At last, the representation of $\mathcal{S}_u$ can be obtained by concatenating the representation ($\mathbf{s}_{i}$) of each item in $\mathcal{S}_u$,
\begin{equation}\label{equ:sequence-representation}
\footnotesize
\mathbf{S}=Concat(\mathbf{s}_{1},\mathbf{s}_{2},\cdots, \mathbf{s}_{n})^T
\end{equation}
which is then fed into DTE for further refinement.

\subsection{Dynamic Transformer Encoder (DTE)}
\subsubsection{\textbf{User-specific Transformer Encoders}}
A fundamental assumption of the transformer architecture is that the transform matrices used in the self-attention layer are shared by all input samples. However, user intentions could be diverse and complex in recommendation scenarios, thus the prediction model should be different for each user to achieve maximal personalization.
Having realized the vital role of user-specific recommendation engines, we hence propose the DTE as described in Fig.~\ref{fig:framework}(b). By performing attention network over the self-attention layer in the transformer encoder, the capacity of the transformer encoder is improved without increasing its depth (blocks of transformer encoders) or width (number of heads). 
An illustration of how we obtain user-specific queries ($\mathbf{Q}_{u}$), user-specific keys ($\mathbf{K}_{u}$), and user-specific values ($\mathbf{V}_{u}$) is presented in Fig.~\ref{fig:dynamic_self_attention}(c). 
Specifically, given the input embedding $\mathbf{S}$ obtained from the IDM, the output of dynamic self-attention $\hat{\mathbf{S}}$ can be described as follows:
\begin{equation}\label{equ:dte}
\footnotesize
\begin{aligned}
&  \hat{\mathbf{S}}= Attention(\mathbf{Q}_{u},\mathbf{K}_{u},\mathbf{V}_{u}) = Softmax(\frac{\mathbf{Q}_{u}\mathbf{K}_{u}^{T}}{\sqrt{d_{k}}})\mathbf{V}_{u} \\
& \text{where }  \mathbf{Q}_{u} = \mathbf{S}\mathbf{W}^{Q}_{u} \text{,  } \mathbf{K}_{u} = \mathbf{S}\mathbf{W}^{K}_{u} \text{, and } \mathbf{V}_{u} = \mathbf{S}\mathbf{W}^{V}_{u}
\end{aligned}
\end{equation}
where $\mathbf{S} \in\mathbb{R}^{n \times d}$ and $\hat{\mathbf{S}} \in\mathbb{R}^{n \times d}$ are the input and output of DTE respectively; $\mathbf{W}^{Q}_{u}$, $\mathbf{W}^{K}_{u}$, and $\mathbf{W}^{V}_{u}$ are learned user-specific transform matrices.
By comparing Eq.~\ref{equ:dte} with Eq.~\ref{equ:self_attention}, it showcases an important merit of our dynamic self-attention module. By assembling the transform matrices before scaled dot-product, it achieves stronger representation power while keeping computational efficiency. We will analyze the efficiency of DTE in Model Complexity and Scalability. 
In what follows, we present how to learn the three user-specific transform matrices $\mathbf{W}^{Q}_{u}$, $\mathbf{W}^{K}_{u}$, and $\mathbf{W}^{V}_{u}$.

\subsubsection{\textbf{Attention over Self-attention}}
Inspired by recent efforts~\cite{iida2019attention,peng2020mixture} which improve multi-head transformer architecture with mixture of experts and attention mechanisms, in this paper, we propose to learn three specialized transform matrices for each user by attentively aggregating multiple transform matrices before the scaled dot-product process: 
\begin{equation}\label{equ:transform_matrices}
\footnotesize
\begin{aligned}
&  \mathbf{W}^{Q}_{u} = a_{1} \mathbf{W}_{1}^{Q} + a_{2} \mathbf{W}_{2}^{Q} + \cdots + a_{t} \mathbf{W}_{t}^{Q}\\
& \mathbf{W}^{K}_{u} = a_{1} \mathbf{W}_{1}^{K} + a_{2} \mathbf{W}_{2}^{K} + \cdots + a_{t} \mathbf{W}_{t}^{K} \\
& \mathbf{W}^{V}_{u} = a_{1} \mathbf{W}_{1}^{V} + a_{2} \mathbf{W}_{2}^{V} + \cdots + a_{t} \mathbf{W}_{t}^{V} 
\end{aligned}
\end{equation}
where $a_t$ is the shared attention weight across the three types of transform matrices, and $t$ is the number of transform matrices for each $\mathbf{Q}_{u}$, $\mathbf{K}_{u}$, and $\mathbf{V}_{u}$.

{In order to learn a meaningful attention weight} $\mathbf{a}=[a_1, a_2,\cdots, a_{t}]$ for each user, we employ an attention network over the self-attention layer in the DTE:
\begin{equation}
\footnotesize
\begin{aligned}
&  \mathbf{a} = Softmax(\mathbf{W}^{A}\mathbf{e} +\mathbf{b}^{A})\\
&  \mathbf{e} = ReLU[\mathbf{W}^{E}\mathbf{ (\mathbf{\overline{p}}_{\mathcal{S}_u} \odot \mathbf{\overline{q}}_{\mathcal{S}_u})}+\mathbf{b}^{E}] 
\end{aligned}
\end{equation}
where $\mathbf{W}^{A},\mathbf{W}^{E}$ are learnable projection matrices; and $\mathbf{b}^{A},\mathbf{b}^{E}$ are bias vectors. 
In order to encode all information in the initial list $\mathcal{S}_{u}$ and learn effective attention weights for each user, we obtain $\mathbf{\overline{p}}_{\mathcal{S}_u}$ and $\mathbf{\overline{q}}_{\mathcal{S}_u}$ by accumulating all latent representations ($\mathbf{p}_{u}$ and $\mathbf{q}_{i}$) learnt via GMF and intention-aware representations ($\mathbf{r}_{i}^{(u)}$ and $\mathbf{r}_{i}^{(i)}$) learnt via IDM,
\begin{equation}
\footnotesize
\begin{aligned}
  &\mathbf{\overline{p}}_{\mathcal{S}_u} = \frac{1}{n} \sum_{i\in\mathcal{S}_{u}} (\mathbf{p}_{u} +\mathbf{r}_{i}^{(u)});
  &\mathbf{\overline{q}}_{\mathcal{S}_u}  = \frac{1}{n} \sum_{i\in\mathcal{S}_{u}} (\mathbf{q}_{i} +\mathbf{r}_{i}^{(i)})
\end{aligned}
\end{equation}
where $n$ is the length of $\mathcal{S}_{u}$.
By estimating $\mathbf{a}$ with information from both implicit feedback data and text review data, the specialized transform matrices can be obtained by Eq.~\ref{equ:transform_matrices}, thus leading to user-specific predictions.

\begin{algorithm}[t]
\footnotesize
\KwIn{$\mathcal{S}_u$, $\mathbf{p}_u$, $\{\mathbf{q}_i|i\in\mathcal{S}_u\}$, $\mathcal{R}_{u}^{e}$, $\{\mathcal{R}_{i}^{e}|i\in\mathcal{S}_u\}$} 
\KwOut{Model parameters $\Theta$, and re-ranked $\mathcal{S}_u$.}
 \For{each user $u \in \mathcal{U}$}{
 \For{each item candidate $i\in\mathcal{S}_{u}$}{
Obtain $\mathbf{r}_{i}^{(u)}$ and $\mathbf{r}_{u}^{(i)}$ with Eqs.~(\ref{equ:co-attention}-\ref{equ:intention-representation})\;
Compute $\mathbf{S}$ for the initial list $\mathcal{S}_{u}$ with Eqs.~(\ref{equ:item-representation}-\ref{equ:sequence-representation})\;
Compute $\hat{y}_{ui}$ with Eq.~\ref{equ:prediction}\;
Calculate $\mathcal{L}$ based on Eq.~\ref{equ:loss}\;
Update $\Theta$ to minimize $\mathcal{L}$, using  $\nabla_{\theta}\mathcal{L}$ ;
 }
 }
\caption{The RAISE Algorithm}
\end{algorithm}
\subsection{Prediction Layer}
In this layer, a new ranking score $\hat{y}_{ui}$ is estimated for each item candidate in the initial list $\mathcal{S}_u$. 
This enables us to generate a final re-ranked list for the target user $u$ by sorting the item candidates according to their new scores from highest.
\begin{equation}\label{equ:prediction}
\footnotesize
\hat{y}_{ui}= Softmax(\mathbf{F}^{(b)} \mathbf{W}^{P} + \mathbf{b}^{P}), i\in \mathcal{S}_u
\end{equation}
{where} $\mathbf{F}^{(b)}$ is the output of $b$ blocks of the DTE. Meanwhile, $\mathbf{W}^P$ and $\mathbf{b}^P$ are transform matrix and bias vector for the prediction layer. We employ a negative log likelihood loss to train our RAISE, as suggested by~\cite{PRM},
\begin{equation}\label{equ:loss}
\footnotesize
\mathcal{L}=-\sum\nolimits_{u\in\mathcal{U}_{tr}} \sum\nolimits_{i\in\mathcal{S}_{u}} y_{ui}\log\hat{y}_{ui}
\end{equation}
where $\mathcal{U}_{tr}$ refers to the set of users in the training set.

\subsection{{Dynamic Self-attention Analysis}}
{ We investigate the properties of our dynamic self-attention  In this section, which is the key of RAISE.
Generally, the self-attention mechanism is usually applied in order to compute the mutual relationships among item candidates in $\mathcal{S}_u$,  whose computational complexity is $\mathcal{O}(n^2d)$.}
Recent studies \cite{pang2020setrank,PRM} propose to capture the mutual relationships from different sub-spaces, multi-head self-attention mechanism can be employed, and its computational complexity is 
$\mathcal{O}(n^2d + nd^2)$, i.e., the extra computational cost is $nd^2$.
In this study, we propose DTE to aggregate transform matrices with attention weights learnt from an attention network based on user intentions.
The key insight is that within reasonable cost of model size, DTE provides user-specific predictions and improves representation capability in an efficient way (low extra computational cost).
Specifically, the extra computational cost of DTE is $\mathcal{O}(td^{2})$ compared with the self-attention mechanism. 
As such, DTE is more computationally efficient than multi-head self-attention 
on the premise that $t\textless n$, which usually holds (the optimal $t$ is no larger than 4, while $n$ is 50 in this study).
Note that DTE takes only one scaled dot-product operation. This means that one can increase the capacity of the prediction model by increasing $t$ with only a small increase in inference cost because each additional parameter requires only 1 additional multiply-add. 

As noted in Table ~\ref{tab:complexity}, our proposed dynamic transformer encoder advances the classic multi-head transformer encoder in following aspects.
(1) Compared to multi-head transformer encoder, DTE provides a more efficient solution to boost the representational capability without increasing the depth (number of DTE blocks $b$) and width (number of heads $h$).
(2) By parameterizing the transform matrices in a dynamic self-attention layer as a linear combination of $t$ experts, DTE is able to perform user-specific (input-dependent) prediction in an efficient manner. 
%
%

\begin{table}[!t]
\centering
\scriptsize
\caption{Per-layer complexity, extra computational cost compared with self-attention (SA) layer, and indication of input-dependent support for different layer types, where $n$ is the sequence length; $d$ is the representation dimension; and $t$ is the number of experts.}\label{tab:complexity}
\vspace{-0.1in}
{
\setlength{\tabcolsep}{7pt}
{
\begin{tabular}{|c|ccc|}
\specialrule{.15em}{.15em}{.15em}
Layer type                & Complexity (per layer)& Extra Cost & Input-dependent \\ \hline
SA           & $\mathcal{O}(n^2d)$                   & N.A              & $\times$             \\ \hline
Multi-head SA & $\mathcal{O}(n^2d+nd^2$ )                   & $nd^2$              & $\times$              \\ \hline
Dynamic SA    & $\mathcal{O}(n^2d)$                    & $td^2$              & $\checkmark$              \\ 
\specialrule{.15em}{.15em}{.15em}
\end{tabular}
}}
\vspace{-0.15in}
\end{table}
\section{Experiments}
In this section, we conduct extensive experiments to answer the following research questions.
\begin{itemize}[leftmargin=*]
    \item \textbf{RQ1:} Is RAISE able to perform competitively to baseline re-ranking solutions? 
    \item \textbf{RQ2:} How does the proposed IDM and DTE affect the performance of RAISE? 
    \item \textbf{RQ3:} Is RAISE able to generate meaningful explanations for item recommendation? 
    \item \textbf{RQ4:} How do the key hyper-parameters affect the performance of RAISE?
{\item \textbf{RQ5:} How does the running cost of RAISE compare with baselines?
    \item \textbf{RQ6:} How do the text reviews affect the performance of re-ranking methods?
    \item \textbf{RQ7:} Can we achieve better results with different co-attention functions and aggregation functions in the IDM module?}
\end{itemize}
\subsection{Experimental Setup}
\textbf{Datasets.}
We conduct experiments on four public datasets from Amazon\footnote{jmcauley.ucsd.edu/data/amazon/} as listed in Table~\ref{tab:statistics}, including `Sports and Outdoors', `Health and Personal Care', `Clothing, Shoes and Jewelry' and `Video Games'.
The selected datasets are widely used in recommendation studies and come from different domains.
After obtaining the initial recommendation lists generated by GMF, we randomly select 80\% users to construct the training set, use 10\% users as the validation set, and leave the remaining 10\% as the testing set for each dataset~\cite{DLCM,PRM}. 
In addition, we follow previous works~\cite{NCF,PRM} to convert explicit rating data to binary implicit feedback. In particular, the label is $1$ if the target user has rated the target item; otherwise 0. 

\begin{table}[t]
\centering
\scriptsize
\caption{Data statistics.}\label{tab:statistics}
\vspace{-0.1in}
{
\setlength{\tabcolsep}{1.2mm}
{
\begin{tabular}{|c|cccc|}
\specialrule{.15em}{.15em}{.15em}
Dataset & \#User & \#Item   & \#Review/\#Rating  & Density \\ \hline
Sports and Outdoors & 35,598  & 18,357   & 296,337
 & 0.05\% \\
Health and Personal Care & 38,609  & 18,534   & 346,355 & 0.05\% \\
Clothing, Shoes and Jewelry & 39,387  & 23,033   & 278,677 & 0.03\% \\

Video Games & 24,303  & 10,672    & 231,780 & 0.09\%\\ 

\specialrule{.15em}{.15em}{.15em}
\end{tabular}
}}
\vspace{-0.15in}
\end{table}

\begin{table*}[!ht]
\renewcommand{\arraystretch}{0.95}
\centering\caption{The performance comparison on all datasets (\%). `Improvement' indicates how much RAISE has improved over the second best method which is underlined. We use $'*'$ to denote statistically significant improvements (paired t-test with $p$-value $<0.05$). Note that "MAP@1" and "NDCG@1" is equal to the "Pre@1", so they are omitted in this table. We run all methods for 5 times and report their average value in this table.} \label{tab:comparison}
\vspace{-0.1in}
\footnotesize
{
\setlength{\tabcolsep}{9pt}
{
\begin{tabular}{cc|ccccccc|}
\specialrule{.15em}{.15em}{.15em}
\multicolumn{1}{|l|}{Datasets}                                                                                   & \multicolumn{1}{c|}{Models}      &
\multicolumn{1}{l}{Pre@1} &
\multicolumn{1}{l}{Pre@5} &
\multicolumn{1}{l}{Pre@10}  & \multicolumn{1}{l}{MAP@5} & \multicolumn{1}{l}{MAP@10}   & \multicolumn{1}{l}{NDCG@5} & \multicolumn{1}{l|}{NDCG@10} \\  
\hline
\multicolumn{1}{|c|}{\multirow{7}{*}{\begin{tabular}[c]{@{}c@{}}Sports\\and \\ Outdoors\end{tabular}}}                   & GMF                          &32.14    & 34.76                                                & 26.73                                                                                                   &46.76  & 48.66  & 42.26 &51.56\\
\multicolumn{1}{|c|}{}                                                                                           & DLCM                         & 55.12    & 41.14                                                & 29.48                                                                                       &63.56 & 60.08   & 52.77 &59.99            \\
\multicolumn{1}{|c|}{}                                             & SetRank$_\text{w/o PE}$                            &71.54&46.65&32.02&75.83&69.64&60.96&66.04\\        \multicolumn{1}{|c|}{}                                         
& {SetRank$_\text{with PE}$}                 &73.17&50.14 &33.90 &77.03&71.90&64.56&69.35           \\
\multicolumn{1}{|c|}{}                                                                                           & PRM                             &\underline{79.44} &\underline{54.83} &\underline{35.39}  &\underline{81.53} &\underline{76.98}  &\underline{70.46} &\underline{74.36}\\
\multicolumn{1}{|c|}{}                                                                                           & RAISE             &$\textbf{85.45}^{*}$ &$\textbf{60.63}^{*}$ &$\textbf{37.77}^{*}$  &$\textbf{86.96}^{*}$ &$\textbf{83.48}^{*}$  &$\textbf{77.84}^{*}$ &$\textbf{80.30}^{*}$     
\\ \cline{2-9}
\multicolumn{1}{|c|}{}                                                                                           & \multicolumn{1}{l|}{Improvement} &7.57\%  & 10.58\%                                             & 6.73\%                                         &6.66\%     & 8.44\%    & 10.47\%   & 7.99\%    \\ \toprule

\multicolumn{1}{|c|}{\multirow{7}{*}{\begin{tabular}[c]{@{}c@{}}Health \\ and \\ Personal \\ Care\end{tabular}}}               
                            & GMF
 &30.65&33.52&26.07&44.66&46.56&40.72&49.65\\
\multicolumn{1}{|c|}{}                                                                                           & DLCM                            &49.23&37.73&27.52&57.70&55.41&48.48&56.17 \\
\multicolumn{1}{|c|}{}                                        & SetRank$_\text{w/o PE}$                            &62.67&43.27&30.18&70.10&65.06&56.25&62.00\\                                   \multicolumn{1}{|c|}{}                            & {SetRank$_\text{with PE}$}                           &68.06&48.07&32.74&73.73&69.30&61.98&67.44\\

\multicolumn{1}{|c|}{}                                                                                           & PRM                             &\underline{73.96} &\underline{52.17} &\underline{34.56}  &\underline{78.04} &\underline{74.15} &\underline{67.61}&\underline{72.01}\\
\multicolumn{1}{|c|}{}                                                                                           & RAISE                         &$\textbf{83.08}^{*}$&$\textbf{59.48}^{*}$ &$\textbf{36.73}^{*}$  &$\textbf{85.53}^{*}$ &$\textbf{82.21}^{*}$  &$\textbf{76.42}^{*}$ &$\textbf{78.99}^{*}$  \\ \cline{2-9}
\multicolumn{1}{|c|}{}                                                                                           & Improvement                    &12.33\%   & 13.95\%                                             & 6.28\%                                                                      &9.60\% & 10.87\%     &13.03\%    &9.69\%                 \\ \toprule

\multicolumn{1}{|c|}{\multirow{7}{*}{\begin{tabular}[c]{@{}c@{}}Clothing, \\ Shoes \\ and \\ Jewelry\end{tabular}}}   
                            & GMF
 &42.83&46.52&33.24&59.69&62.42&57.96&67.19\\
\multicolumn{1}{|c|}{}                                                                                           & DLCM     &68.06&50.79&34.65&73.40&71.10&67.02&74.32\\
\multicolumn{1}{|c|}{}                                      & SetRank$_\text{w/o PE}$                            &79.08&54.60&35.73&81.95&77.10&72.29&77.60\\      
\multicolumn{1}{|c|}{} 
& {SetRank$_\text{with PE}$}           &83.37&59.08&38.28&84.71&80.66&77.22&82.18\\

\multicolumn{1}{|c|}{}                                                                                           & PRM      &\underline{92.06}&\underline{65.94}&\underline{40.35}&\underline{92.15}&\underline{89.08}&\underline{86.69}&\underline{89.00}\\
\multicolumn{1}{|c|}{}                                                                                           & RAISE      &$\textbf{95.30}^{*}$ &$\textbf{70.39}^{*}$ &$\textbf{41.66}^{*}$  &$\textbf{95.32}^{*}$ &$\textbf{93.25}^{*}$  &$\textbf{91.33}^{*}$ &$\textbf{92.38}^{*}$\\ \cline{2-9}
\multicolumn{1}{|c|}{}                                                                                           & \multicolumn{1}{l|}{Improvement} &3.52\% & 6.75\%                                              & 3.25\%                                                                                                    &3.44\%     & 4.68\%    & 5.35\% & 3.41\%    \\ 
\toprule
\multicolumn{1}{|c|}{\multirow{7}{*}{\begin{tabular}[c]{@{}c@{}}Video \\ Games \end{tabular}}} & GMF           &40.80&43.66&34.23&57.23&57.86&50.39&58.63      \\
\multicolumn{1}{|c|}{}                                                                                           & DLCM     &58.82&47.09&35.33&66.77&64.00&56.68&63.86 \\
\multicolumn{1}{|c|}{}                                      & SetRank$_\text{w/o PE}$                            &69.47&50.16&36.19&76.06&70.55&61.93&66.76\\  
\multicolumn{1}{|c|}{} & {SetRank$_\text{with PE}$} &71.93&54.11&38.53&77.37&72.78&65.90&70.57\\
\multicolumn{1}{|c|}{}                                                                                           & PRM      &\underline{74.65}&\underline{55.76}&\underline{39.03}&\underline{79.58}&\underline{75.14}&\underline{68.07}&\underline{72.15}\\
\multicolumn{1}{|c|}{}                                                                                           & RAISE    &$\textbf{81.35}^{*}$ &$\textbf{59.43}^{*}$ &$\textbf{40.68}^{*}$  &$\textbf{83.99}^{*}$ &$\textbf{79.33}^{*}$  &$\textbf{73.16}^{*}$ &$\textbf{76.30}^{*}$\\ \cline{2-9}
\multicolumn{1}{|c|}{}                                                                                           & \multicolumn{1}{l|}{Improvement} &8.98\% & 6.58\%                                              & 4.23\%                                                                                                &5.54\%     & 5.58\%    & 7.48\% & 5.75\%    \\ 
\specialrule{.15em}{.15em}{.15em}
\end{tabular}
}}
\vspace{-0.15in}
\end{table*}

\begin{table}[!t]
\centering
\scriptsize
\caption{The effects of key components of RAISE (\%).}\label{tab:ablation2}
\vspace{-0.1in}
{
\setlength{\tabcolsep}{4pt}
{
\begin{tabular}{|c|c|cc|cc|cc|}
\specialrule{.15em}{.15em}{.15em}
\multirow{2}{*}{Dataset} &\multirow{2}{*}{Model} &\multicolumn{2}{c|}{Pre} &\multicolumn{2}{c|}{MAP} &\multicolumn{2}{c|}{NDCG} \\\cline{3-8}
& &@5 &@10 &@5 &@10 &@5 &@10 \\ \hline
                                                                           & RAISE                                                       & \textbf{60.63}                               & \textbf{37.77}& \textbf{86.96}             & \textbf{83.48}                                       & \textbf{77.84}     & \textbf{80.30}                      \\ 
                                                                           & RAISE$_\text{w/o IDM}$                                                      & 59.70                                             & 37.44                     & 86.88                                  & 82.76  & 76.85 	&79.40
                       \\  
                                                                           & \begin{tabular}[c]{@{}l@{}}RAISE$_\text{w/o DTE}$ \\ \end{tabular} & 59.18                                                   & 37.22                   &86.14                       & 82.03          & 76.23    	 &78.94
                   \\ 
\multirow{-4}{*}{\begin{tabular}[c]{@{}c@{}}Sports \\and\\ Outdoors\end{tabular}} & RAISE$_\text{w/o Both}$                                                           & 58.50                                                 & 36.78 &86.00                                                  & 81.83 & 75.55
	&78.28
\\ \toprule
                                                                           & RAISE                                                       & \textbf{59.48}                                                 & \textbf{36.73}   & \textbf{85.53}                                        & \textbf{82.21}         & \textbf{76.42}      & \textbf{78.99}                       \\  
                                                                           & RAISE$_\text{w/o IDM}$                                                      & 58.34                                                 & 36.12                       &84.08                            & 81.02  & 75.06       	&77.92
                \\  
                                                                           & \begin{tabular}[c]{@{}l@{}}RAISE$_\text{w/o DTE}$ \\ \end{tabular} & 58.54                                                 & 36.57                     &84.78                        & 81.24        & 75.38     	&78.26
                  \\ 
\multirow{-4}{*}{\begin{tabular}[c]{@{}c@{}}Health \\and\\Personal\\Care\end{tabular}} &  RAISE$_\text{w/o Both}$                                                           & 53.10                                                & 34.38                                              &79.93        & 76.03 & 68.95 	&73.14
                     
\\ \toprule
                                                                           & RAISE                                                       & \textbf{70.39}                                                 & \textbf{41.66}    & \textbf{95.32}                                        & \textbf{93.25}         & \textbf{91.33}      & \textbf{92.38}                       \\  
                                                                           & RAISE$_\text{w/o IDM}$                                                      & 69.10                                                 & 40.86                       &94.81                            & 92.59  & 90.19       	&91.53
                \\  
                                                                           & \begin{tabular}[c]{@{}l@{}}RAISE$_\text{w/o DTE}$ \\ \end{tabular} & 70.12                                                 & 41.56                     &95.26                        & 93.12        & 91.07     	&92.37
                  \\ 
\multirow{-4}{*}{\begin{tabular}[c]{@{}c@{}}Clothing, \\ Shoes\\and\\Jewelry\end{tabular}} &  RAISE$_\text{w/o Both}$                                                           & 67.68                                                & 40.58                                              &93.36        & 91.06 & 88.50 	&90.51

\\ \toprule
                                                                           & RAISE                                                       & \textbf{59.43}                                                 & \textbf{40.68}    & \textbf{83.99}                                        & \textbf{79.33}         & \textbf{73.16}      & \textbf{76.30}                       \\  
                                                                           & RAISE$_\text{w/o IDM}$                                                      & 57.43                                                 & 39.65                       &82.06                            & 77.51  & 70.66       	&74.08
                \\  
                                                                           & \begin{tabular}[c]{@{}l@{}}RAISE$_\text{w/o DTE}$ \\ \end{tabular} & 58.77                                                 & 40.39                     &83.52                        & 78.79        & 72.42     	&75.49
                  \\ 
    \multirow{-4}{*}{\begin{tabular}[c]{@{}c@{}}Video \\Games \end{tabular}} &  RAISE$_\text{w/o Both}$                                                           & 56.39                                                & 39.39                                              &81.14        & 76.24 & 69.43 	&73.30
                     \\ 
                     \specialrule{.15em}{.15em}{.15em}
\end{tabular}
}}
\vspace{-0.15in}
\end{table}

\smallskip\noindent\textbf{Comparing Methods.}
We compare with the following state-of-the-art counterparts.
\begin{itemize}[leftmargin=*]
    \item \textbf{GMF}~\cite{NCF} generalizes the matrix factorization model in a non-linear manner, which is a widely adopted baseline for recommendation tasks. Note that GMF is the prior global ranking model of all re-ranking baselines in this paper.
    \item \textbf{DLCM}~\cite{DLCM} is a classic re-ranking model, which encodes item candidates in the initial list sequentially with GRU. 
    \item \textbf{PRM}~\cite{PRM} {is constructed based on the transformer architecture. It employs a pre-trained model to generate personalized vectors for candidate items, which are then fed into the transformer network to refine initial item lists together with the latent representations learnt from GMF. }
    \item \textbf{SetRank}~\cite{pang2020setrank} {re-ranks items with the multi-head self-attention mechanism and an attentive loss function.} We implement both the SetRank with and without positional embeddings, which are denoted as SetRank$_\text{with PE}$ and SetRank$_\text{w/o PE}$, respectively.
\end{itemize}

Note that the source code of Seq2Slate~\cite{bello2018seq2slate} is not released and our re-implemented version performs unsatisfactorily. Meanwhile, we also find that the performance of IRGPR~\cite{liu2020personalized}\footnote{https://github.com/wwliu555/IRGPR} is poor in our experimental setting, although we have carefully tuned its parameters. Therefore, we omit the comparison with Seq2Slate and {IRGPR} in this paper.


\smallskip\noindent\textbf{Training Details.}
For a fair comparison, we set $d=32$ and $n=50$ for all re-ranking methods. 
We follow the configuration presented in \cite{NCF,DLCM,PRM,pang2020setrank}, and all the baselines are trained until convergence.
For our proposed RAISE, we tune the number of hidden layers from 1 to 4 for MLP structures. {The maximum number of reviews} $\textit{l}_{u}$ and $\textit{l}_{i}$ are set to 20.
{Instead of constructing and fine-tuning for an end-to-end NLP model to obtain the word representations for text information, we obtain pre-trained language model word representations from BERT's pre-trained model in this work since many NLP tasks are benefit from BERT to get the SOTA \cite{devlin2018bert}.} 
The learning rate is selected from $\{1e-1,1e-2,1e-3,1e-4\}$; batch size is chosen from $\{256, 512, 1024\}$ and dropout rate varies in the range of $[0.1, 0.5]$ stepped by 0.1. 
Moreover, the number of transform matrices $t$ and the number of DTE blocks $b$ are searched from $\{1,2,4,8,10\}$, and $\{1,2,3,5,8,10\}$.
Our model is implemented with Pytorch \footnote{pytorch.org/}, optimized with  Adam~\cite{kingma2014adam}, and trained on one Nvidia TITAN Xp GPU with 12 GB memory associated with Intel Exon CPU E5-2630 v4@2.20GHz. 

\smallskip\noindent\textbf{Evaluation Metrics.} We adopt the same evaluation metrics with PRM~\cite{PRM} to evaluate the performance of all methods: Precision (Pre@$k$) and Mean Average Precision (MAP$@k$), where $k$ is the length of the recommendation list. \textit{Precision} evaluates the fraction of correct recommendations in recommendation lists for all users, and \textit{MAP} computes the mean average precision of all ranked lists cut off by $k$. 
In addition, we also evaluated the recommendation performance by normalized discounted cumulative gain (NDCG@$k$) which takes the
position of correct recommendations into account \cite{NCF,lin2020comet}.
Note that higher metric values indicate a better recommendation performance.

\subsection{Experimental Results}

\subsubsection{\textbf{Performance Comparison (RQ1)}}
Table~\ref{tab:comparison} presents the overall performance of the proposed RAISE and comparing methods on four Amazon datasets. As a whole, all re-ranking methods are able to refine the initial recommendation list generated by the global ranking model GMF. This verifies the effectiveness of modeling {inter-item} patterns for re-ranking tasks. Moreover, we can observe that RAISE performs better than the other three state-of-the-art re-ranking models: DLCM, {SetRank} and PRM. 
In particular, RAISE achieves up to 13.95\% relative improvement w.r.t. Pre@5, 12.30\% relative improvement w.r.t. MAP@5, and 9.60\% relative improvement w.r.t. NDCG@5 on `Health and Personal Care' dataset, compared to the second best re-ranking algorithm. 
We also observe that RAISE gains slightly less improvement on the 'Clothing, Shoes and Jewelry' dataset. A possible explanation could be the difficulty of model training due to the extremely sparse data, especially for RAISE which refines recommendations with review information, as we can see the review density of this dataset is lower than 0.05\% as shown in Table~\ref{tab:statistics}.  
Another interesting finding is that considering the positional embeddings for SetRank leads to better results on the four datasets.

\begin{figure*}[!htbp]
\centering
    \subfigure[Sports and Outdoors]{
    \includegraphics[width=0.22\textwidth]{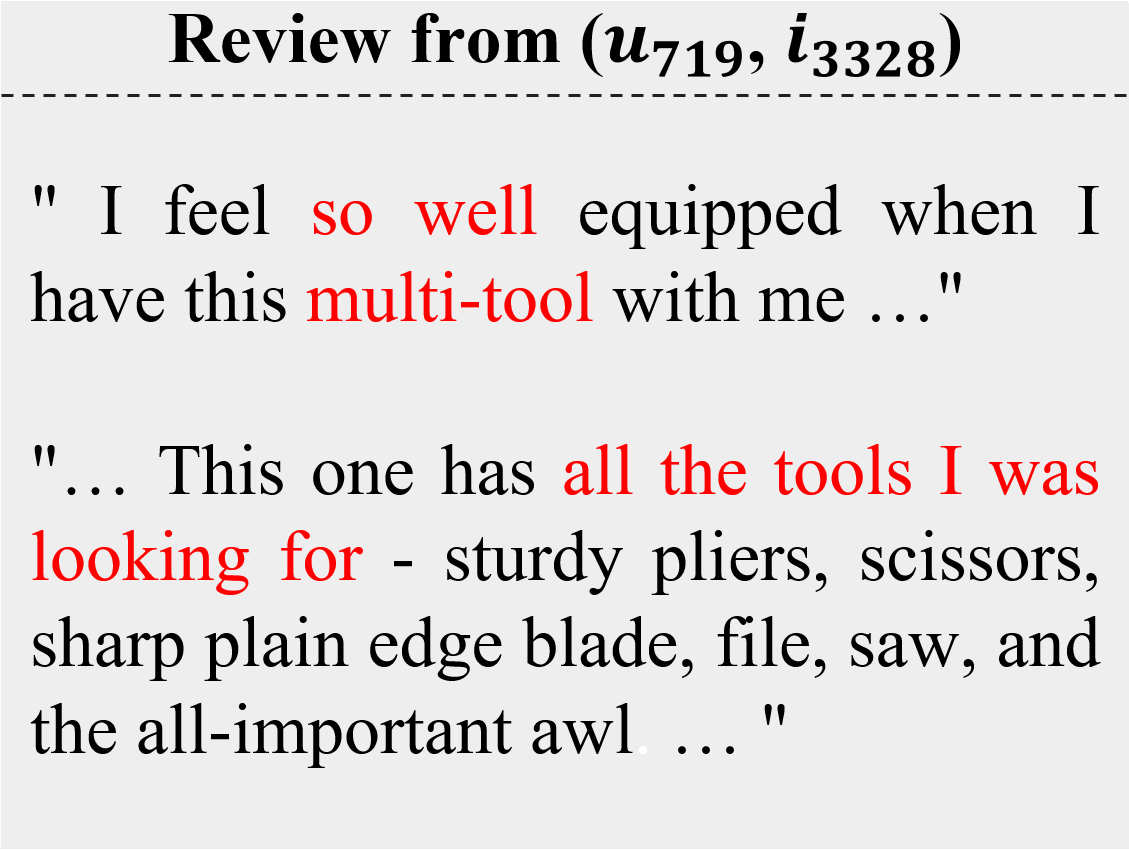}}
   \hspace{0.1in}  
    \subfigure[Health and Personal Care]{
    \includegraphics[width=0.22\textwidth]{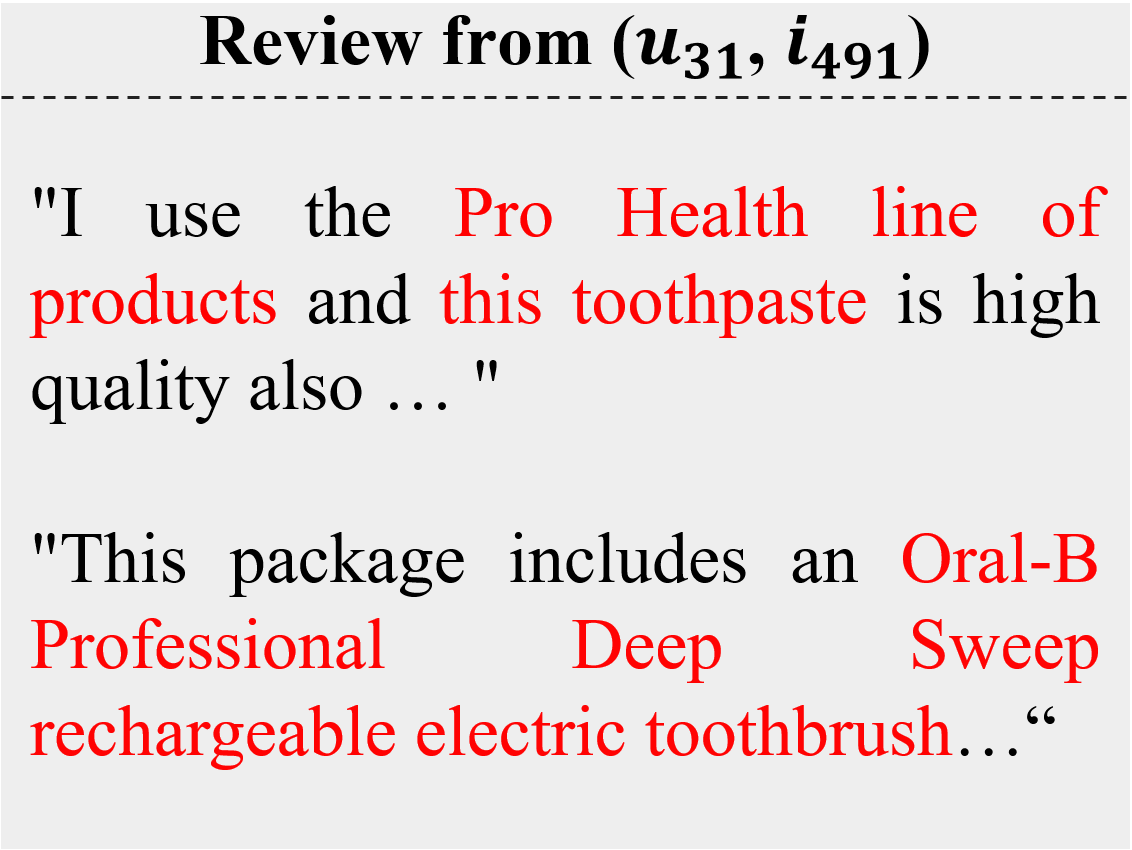}}
    \hspace{0.1in}
    \subfigure[Clothing, Shoes and Jewelry]{
    \includegraphics[width=0.22\textwidth]{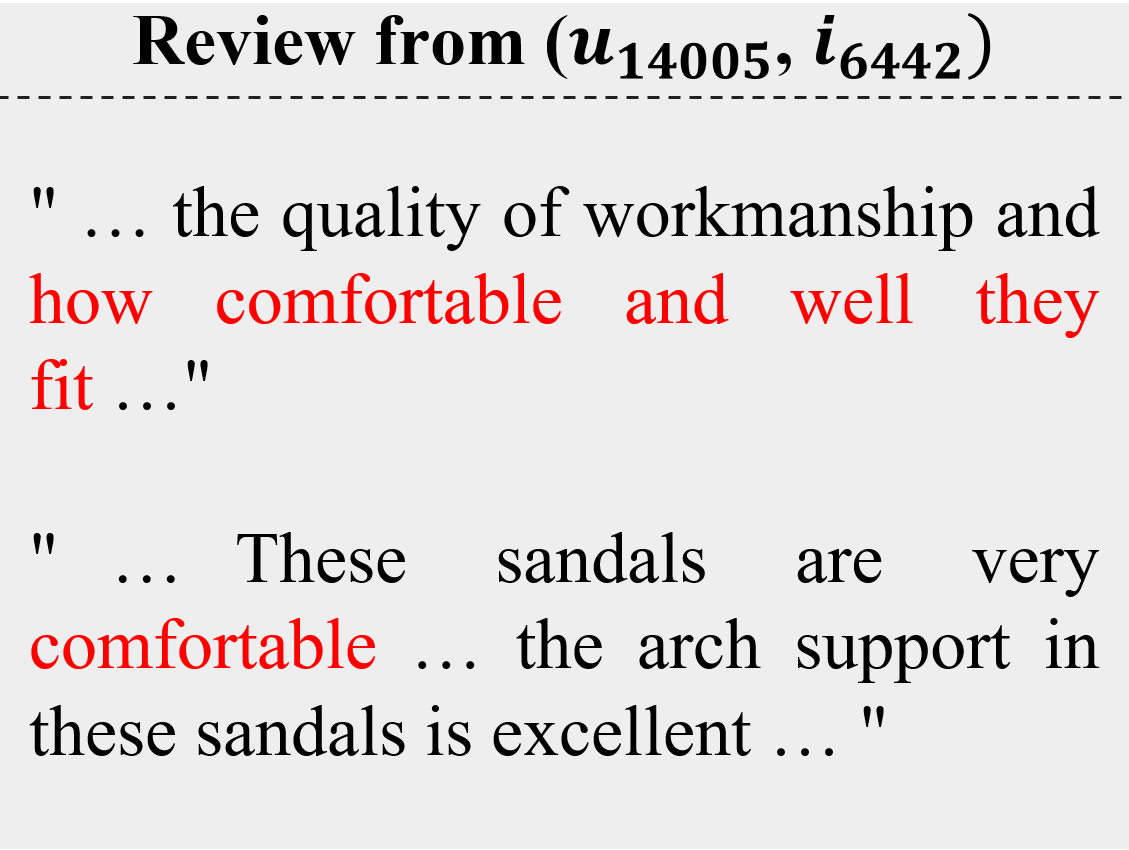}}
    \hspace{0.1in} 
    \subfigure[Video Games]{
    \includegraphics[width=0.22\textwidth]{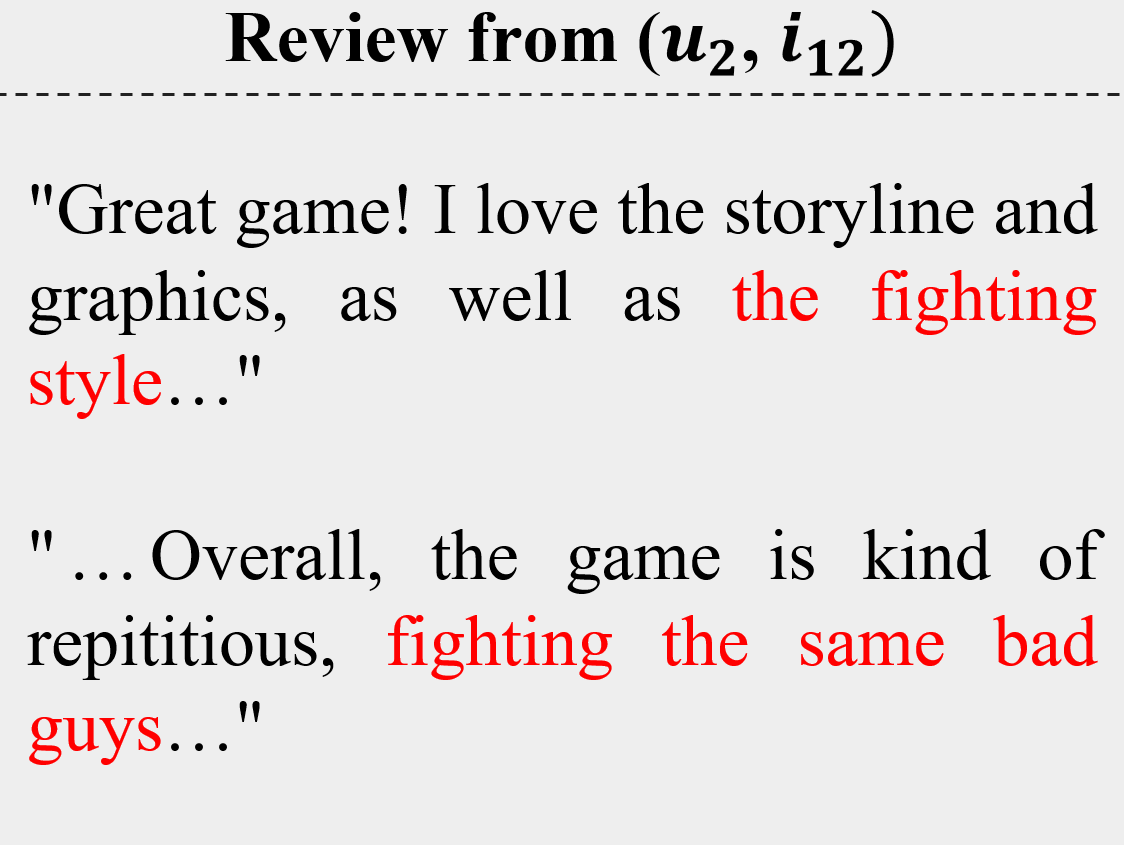}}
    
    \subfigure[Sports and Outdoors]{
    \includegraphics[width=0.22\textwidth]{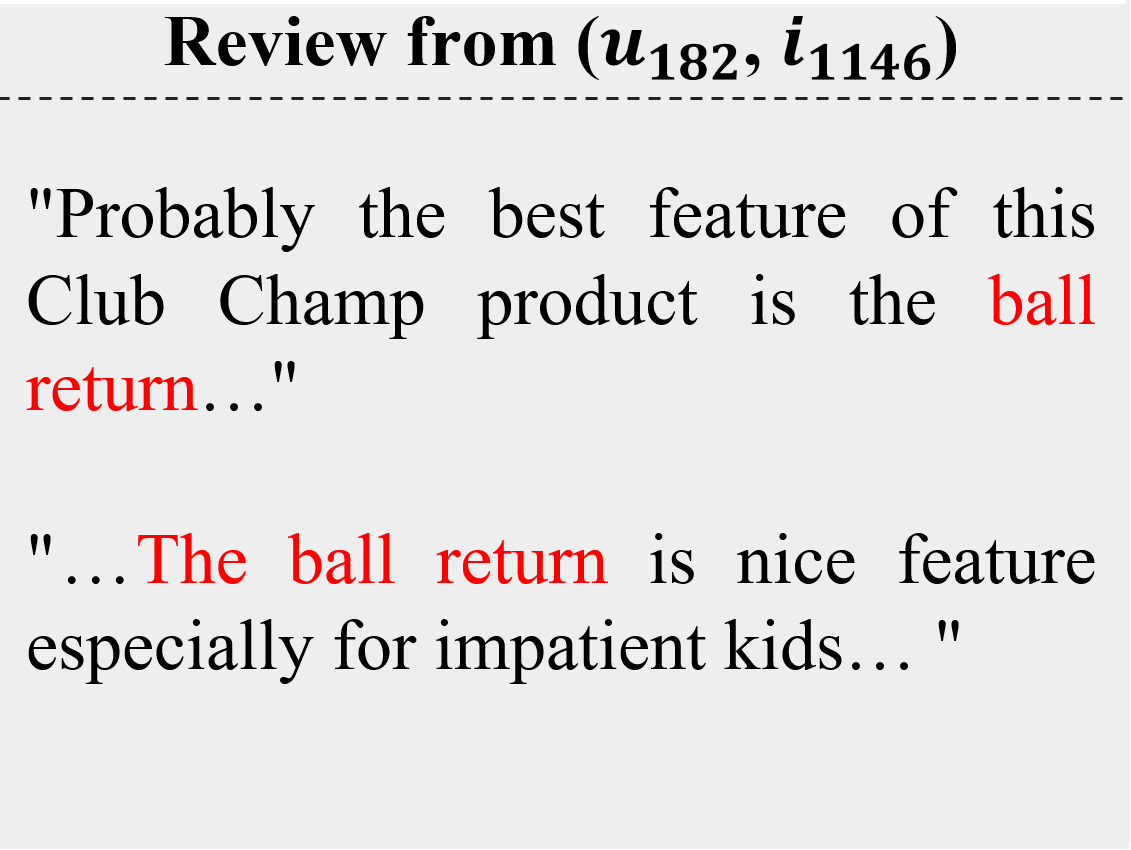}}
    \hspace{0.1in}  
    \subfigure[Health and Personal Care]{
    \includegraphics[width=0.22\textwidth]{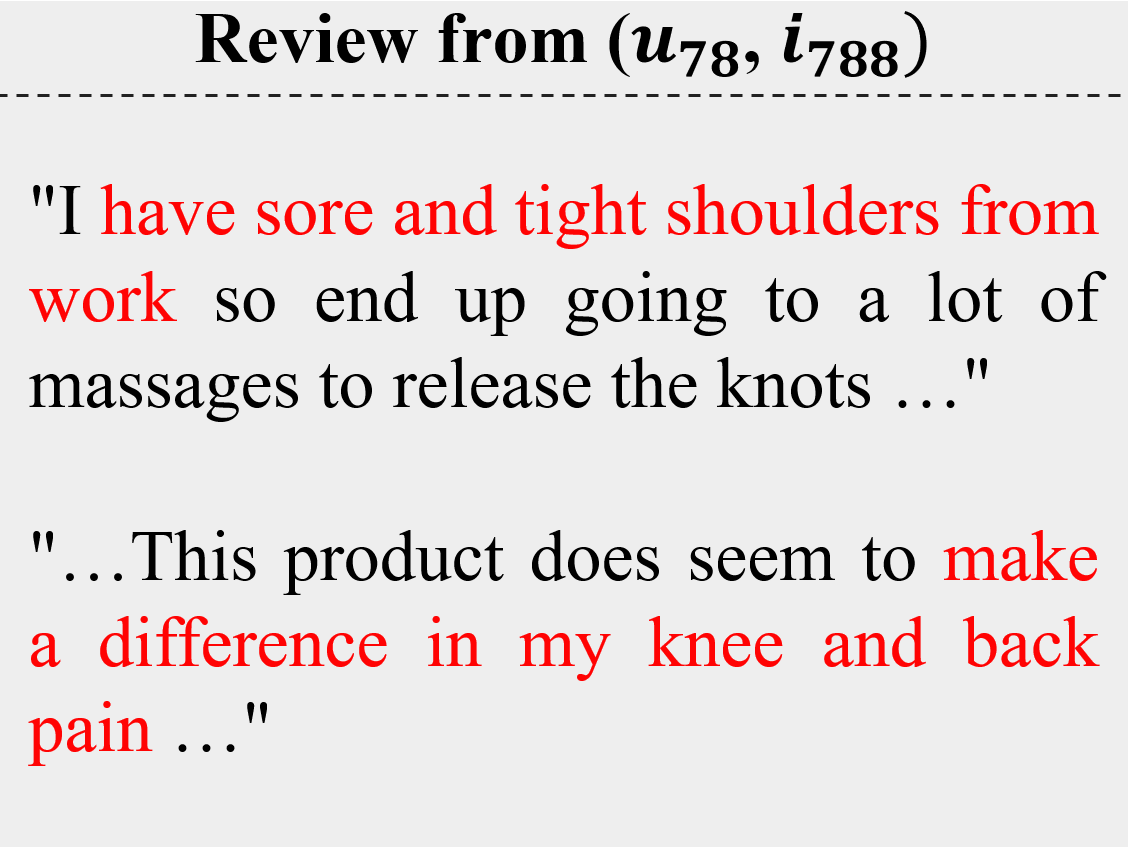}}
    \hspace{0.1in}
    \subfigure[Clothing, Shoes and Jewelry]{
    \includegraphics[width=0.22\textwidth]{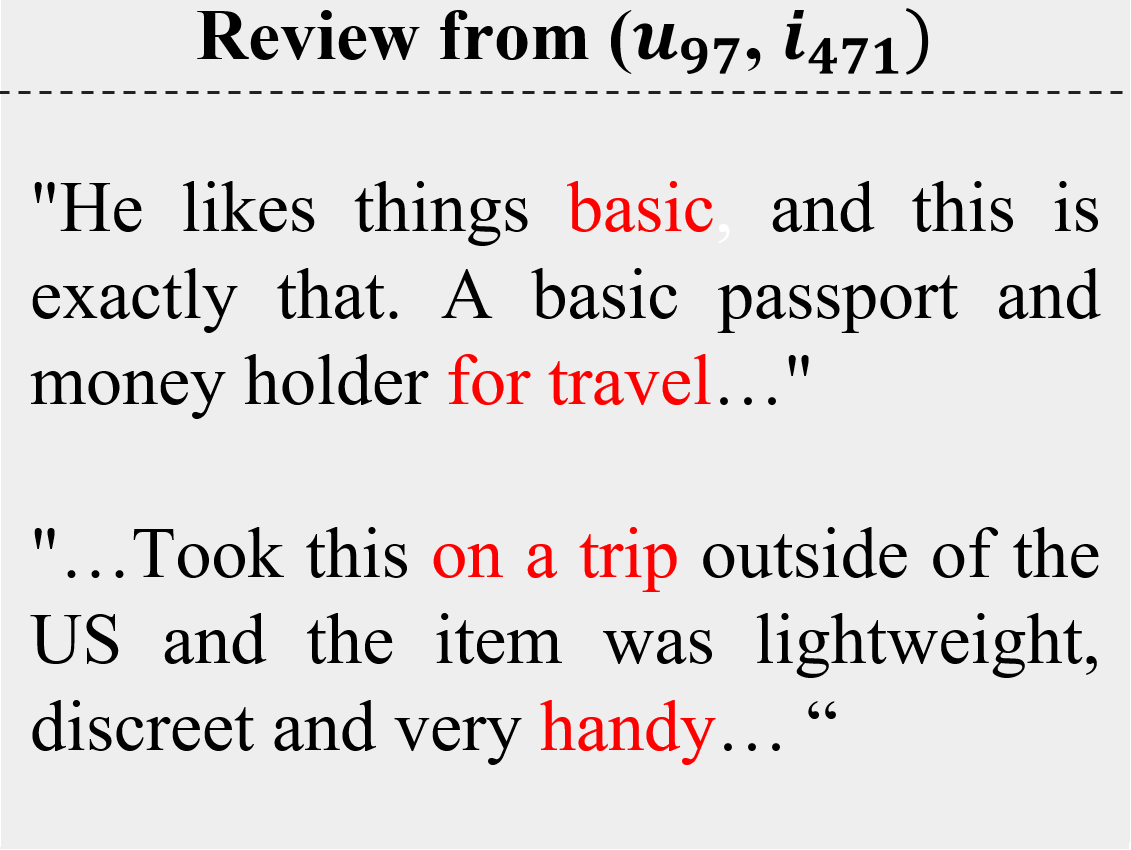}}
    \hspace{0.1in} 
    \subfigure[Video Games]{
    \includegraphics[width=0.22\textwidth]{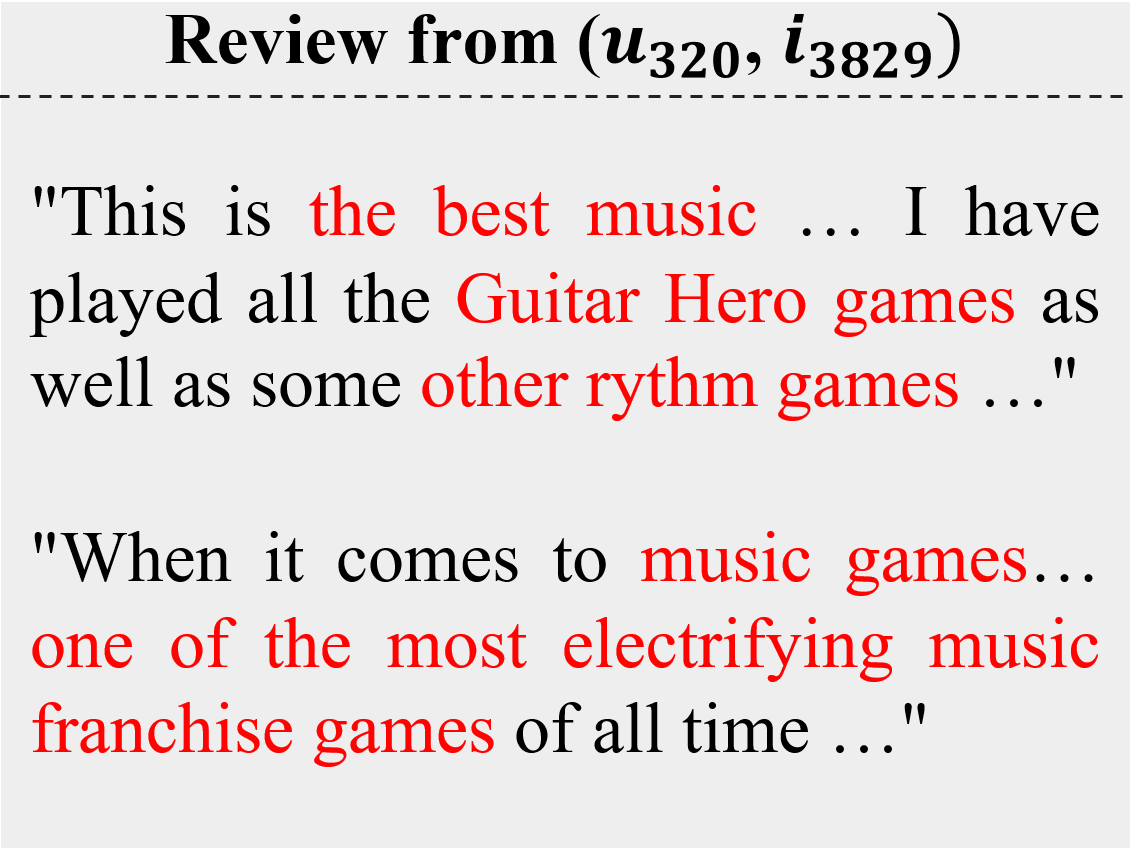}}
  \vspace{-0.1in}
  \caption{Selected user-item review pairs with high matching scores from the four datasets.
  }\label{fig:explanation} 
  \vspace{-0.1in}
\end{figure*}

\begin{figure}[!ht]
\centering
    \subfigure[Sports and Outdoors]{
    \includegraphics[width=0.23\textwidth]{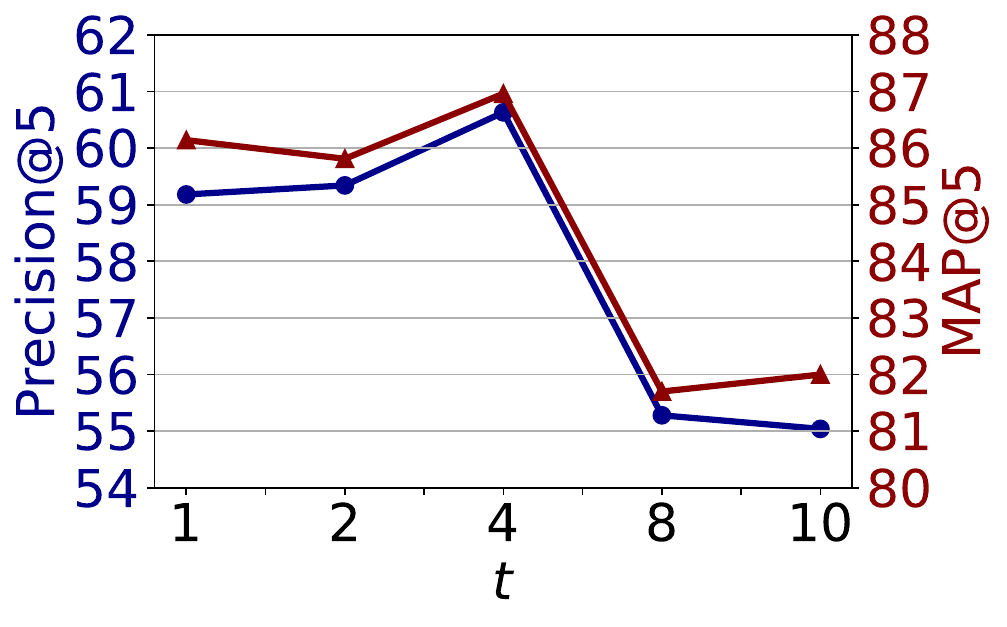}}
    \subfigure[Video Games]{
    \includegraphics[width=0.23\textwidth]{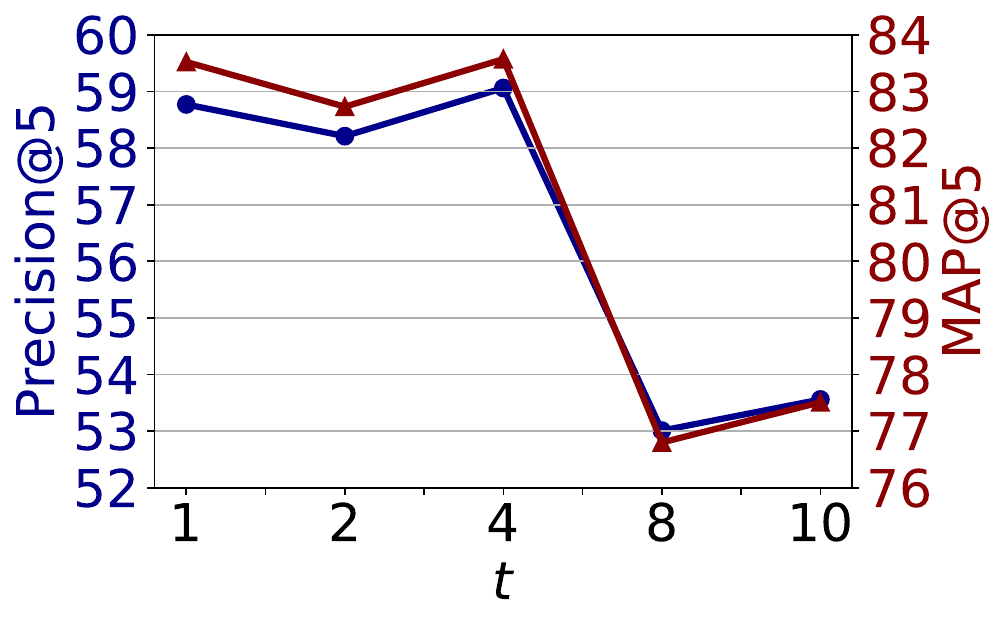}}
    \vspace{-0.1in}
    \caption{Impact of $t$ at Pre@5(\%) and MAP@5(\%).}
  \label{fig:parameter-t} 
  \vspace{-0.1in}
\end{figure}
\begin{figure}[!ht]
\centering
    \subfigure[Sports and Outdoors]{
    \includegraphics[width=0.23\textwidth]{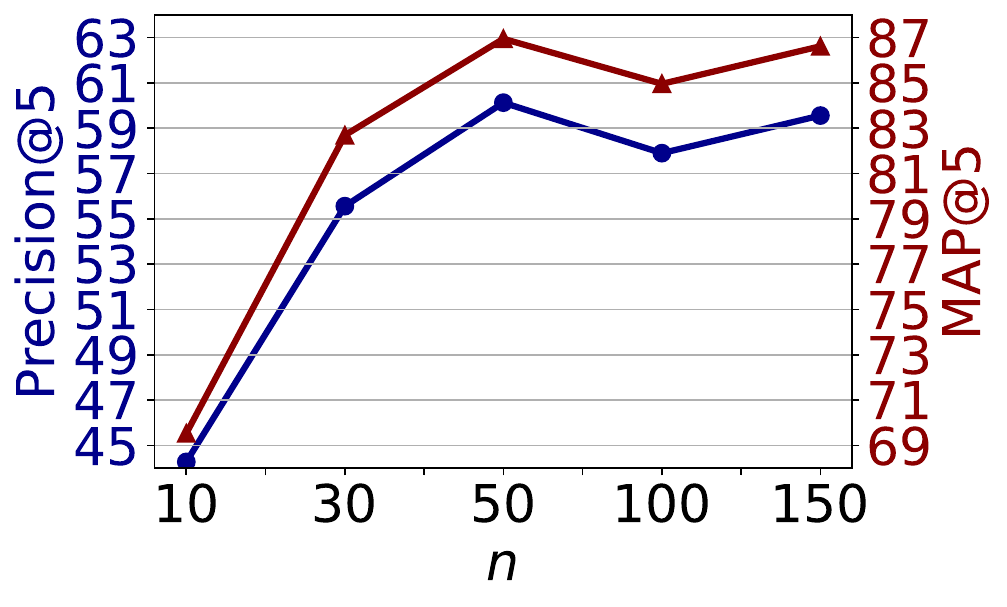}}
    \subfigure[Video Games]{
    \includegraphics[width=0.23\textwidth]{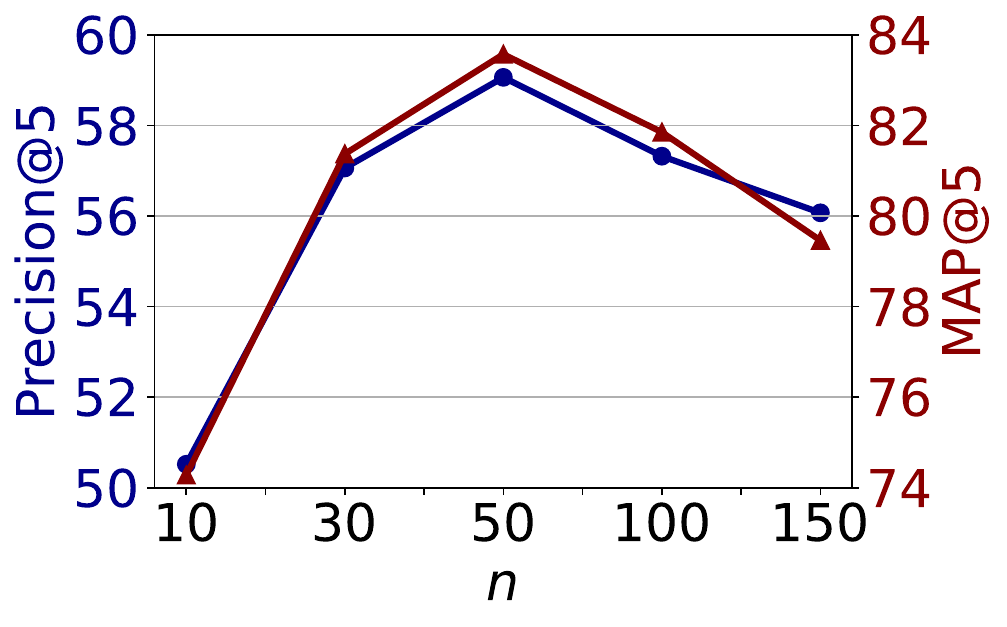}}
    \vspace{-0.1in}
    \caption{Impact of $n$ at Pre@5(\%) and MAP@5(\%).}
  \label{fig:parameter-n} 
  \vspace{-0.1in}
\end{figure}
\begin{figure}[t]
\centering
    \subfigure[Sports and Outdoors]{
    \includegraphics[width=0.23\textwidth]{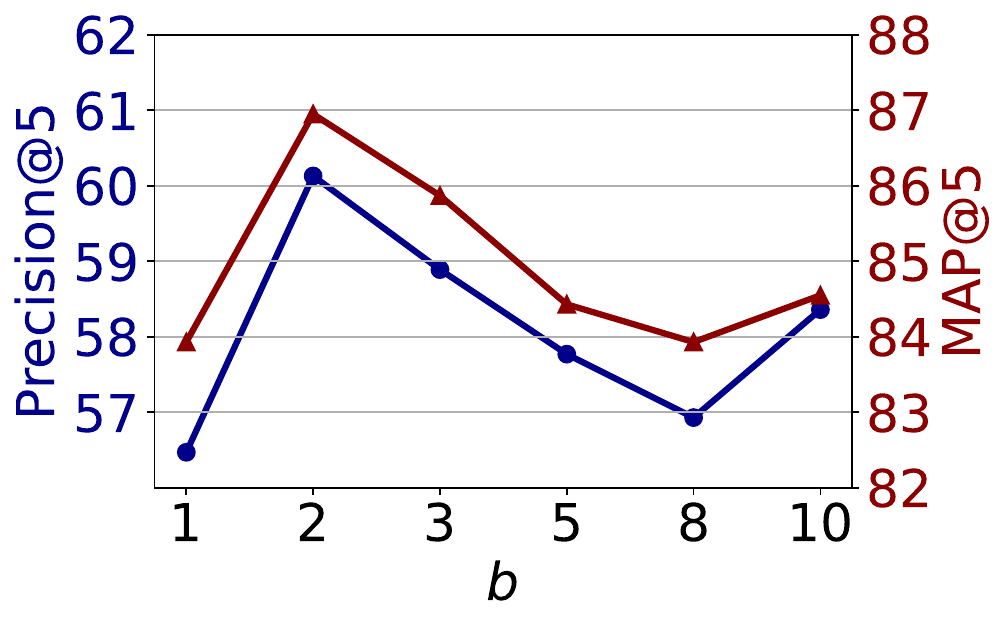}}
    \subfigure[Video Games]{
    \includegraphics[width=0.23\textwidth]{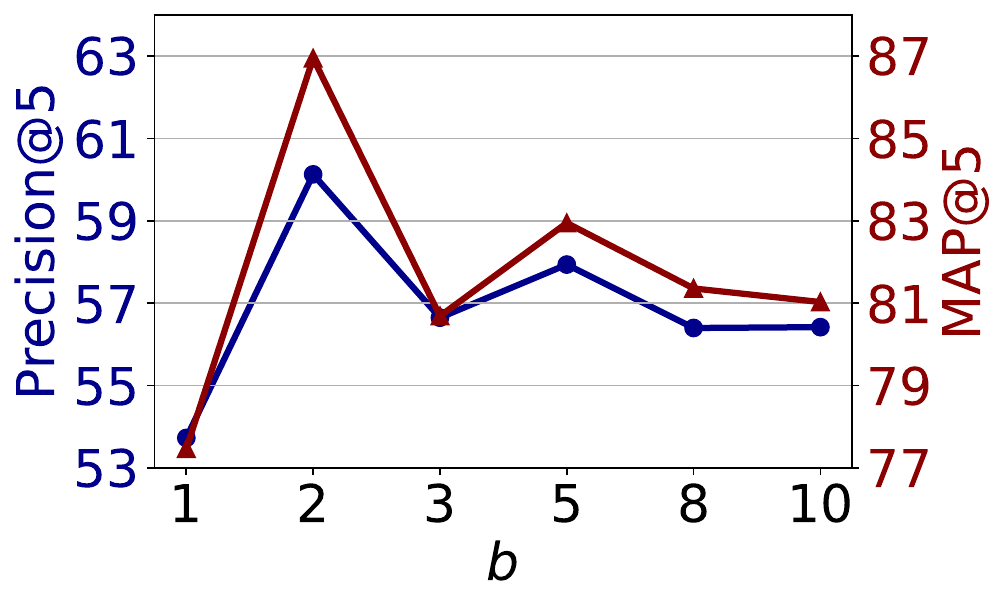}}
    \vspace{-0.1in}
    \caption{Impact of $b$ at Pre@5(\%) and MAP@5(\%).}
  \label{fig:parameter-b} 
  \vspace{-0.1in}
\end{figure}

\subsubsection{\textbf{Effects of IDM and DTE (RQ2)}}
To investigate the effectiveness of two key components of RAISE, we compare RAISE with its three variants listed as follows: 
(1) RAISE$_\text{w/o IDM}$: removing the intention-aware representations learnt via IDM in Eq.~\ref{equ:item-representation};
(2) RAISE$_\text{w/o DTE}$: replacing DTEs with normal transformer encoders as shown in Eq.~\ref{equ:self_attention};
(3) RAISE$_\text{w/o Both}$: removing both IDM and DTEs from RAISE simultaneously. The results are reported in Table~\ref{tab:ablation2}.
By analyzing the performance comparison, we have the following observations.
(1) Being equipped with IDM, our proposed RAISE is able to {effectively} learn useful patterns from review information and further enrich the representations of users and items. 
(2) By employing an attention network over the self-attention layer, the proposed DTE performs better than the normal transformer encoder under our setting, showing its stronger representational capability without increasing the depth or width.
(3) The combination of IDM and DTE brings the largest improvements, which showcases that RAISE is able to generate more personalized recommendations by seamlessly accommodating the learnt latent user intentions extracted from the text reviews.

One may wonder why performance gaps exist among RAISE$_\text{w/o Both}$, SetRank, and PRM methods since they are all based on the transformer architectures.
This could come from the following three reasons.
First, the input data is different, which results in the different performance. 
In particular, RAISE improves the personalization of the recommendation engine by empowering it to be input-dependent, while the PRM model employing a pre-trained model to generate personalized vectors for candidate items, which are learned from implicit feedback.
As mentioned in their paper, such personalized vectors encode users' generic preferences and could be beneficial to the predictions.
They are then fed into the transformer architecture together with item latent representations $q_i$.
As such, the input of PRM is not the same as ours, because RAISE$_\text{w/o Both}$ is trained with both user representations $p_u$ and item representations $q_i$ in an end-to-end manner, rather than encoding user information via a pre-trained model.   
In summary, SetRank and RAISE$_\text{w/o both}$ take the latent representation of users $p_u$ as input, whereas PRM does not.
Second, the range of hyper-parameters are different. 
As mentioned in the section 5.1, in order to reproduce the results of comparing methods and make a fair comparison, we follow the same configurations presented in their paper, including the search spaces of hyper-parameters, which could make a considerable impact on the model performance. 
Third, the choice of loss functions could be influential to model performance.
To be specific, SetRank adopts an attentive loss function, which is different from the negative log likelihood loss used in PRM and RAISE.

\subsubsection{\textbf{Interpretability of RAISE (RQ3)}}
We now discuss another important property of our proposed RAISE: the interpretability.
To this end, we randomly pick eight user-item review pairs with high matching scores among the testing set of the four datasets and highlight the similar intention(s) between user and item reviews in Fig.~\ref{fig:explanation}. 
Taking Fig.~\ref{fig:explanation}(a) as an example, user 719 wrote about "I feel so well equipped when I have this multi-tool with me", and then the co-attention mechanism assigns higher matching score to the item review that mentioned "This one has all the tools I was looking for". 
We find that, even there are very few words that occur in common between two reviews, the selected review pairs are identified to be consistent with regard to some high-level concepts. 
This indicates it is a practical way to find meaningful item reviews for the target user. By distinguishing the importance of reviews between two input sequences $\mathcal{R}_{u}$ and $\mathcal{R}_{i}$, RAISE is capable of boosting the interpretability of recommendation engines.

\subsubsection{\textbf{Hyper-parameter Sensitivity Analysis (RQ4)}}
In this section, we examine how the number of transform matrices ($t$), the length of the initial recommendation list ($n$) and {the number of DTE blocks} ($b$) affect the performance of RAISE. The results are depicted in Figs.~(\ref{fig:parameter-t}-\ref{fig:parameter-b}), and similar trends can be observed with $k=\{10, 20\}$ as well as the rest two datasets.  

From Fig.~\ref{fig:parameter-t}, we can see that {a small number} of transform matrices ($t\leq 4$) is sufficient to distinguish the main intentions of users. This indicates our DTE does not actually add much space complexity and verifies our assumption that user behavior is driven by multiple intentions with different contributions. 
From Fig.~\ref{fig:parameter-n}, we notice that the performance of RAISE initially improves with the increase of n ($n \leq 50$), and the performance begins to decrease after a certain point ($n \textgreater 50$) on both datasets. Intuitively, a longer initial list contains more item candidates and richer {inter-item} patterns, which is beneficial to re-ranking models. However, since RAISE is empowered by the transformer architecture, it may not be a good choice to blindly increase the length of initial list, {as the transformer architecture is quadratic to} $n$ (see Model Complexity and Scalability). Besides, a long initial list could be hard for the transformer architecture to model effective mutual relationships.
Furthermore, the results on {the number of DTE blocks} are plotted in Fig.~\ref{fig:parameter-b}, indicating that only one or two DTE {blocks} are usually sufficient for RAISE to achieve its optimal performance.

\begin{table}[!t]
\centering
\scriptsize
\caption{{Training time of RAISE and comparing methods (hours:minutes:seconds). We follow the default setting of DLCM and train it for 10,000 iterations. As for PRM, SetRank RAISE and RAISE$_\text{w/o IDM}$, they are trained for 100 epoch respectively.}}\label{tab:running time}
\vspace{-0.1in}
{
\setlength{\tabcolsep}{4pt}
{
\begin{tabular}{|c|c|c|c|c|c|}

\specialrule{.15em}{.15em}{.15em}
\multirow{2}{*}{Dataset}  &\multicolumn{5}{c|}{Model}  \\\cline{2-6}
 &DLCM &PRM &SetRank &RAISE &RAISE$_\text{w/o IDM}$   \\ \hline
                                                            \multirow{-1}{*}{\begin{tabular}[c]{@{}c@{}}Sports and \\Outdoors\end{tabular}}                                                                    & \multirow{2}{*}{04:59:51}                               & \multirow{2}{*}{00:10:54}& \multirow{2}{*}{00:26:09}             & \multirow{2}{*}{04:22:58}  & \multirow{2}{*}{00:24:08}               \\                    
                                         &&&&&                   
\\ \toprule
                                                            \multirow{-1}{*}{\begin{tabular}[c]{@{}c@{}}Health and\\Personal Care\end{tabular}}                                                       & \multirow{2}{*}{05:35:59}                                         & \multirow{2}{*}{00:09:13}  & \multirow{2}{*}{00:26:48}                                       & \multirow{2}{*}{05:01:59}      & \multirow{2}{*}{00:24:59}                     \\  
                                        &&&&&                         
\\ \toprule
                                                            \multirow{-1}{*}{\begin{tabular}[c]{@{}c@{}}Clothing, Shoes\\and Jewelry\end{tabular}}                                                        & \multirow{2}{*}{05:36:45}                                               & \multirow{2}{*}{00:11:57}   & \multirow{2}{*}{00:27:29}                                      & \multirow{2}{*}{05:14:56}          & \multirow{2}{*}{00:19:14}                   \\  
                                    &&&&&
                     
\\ \toprule
                                                                \multirow{-1}{*}{\begin{tabular}[c]{@{}c@{}}Video \\Games \end{tabular}}                                                                  & \multirow{2}{*}{04: 59: 51}                                              & \multirow{2}{*}{00:06:14}    & \multirow{2}{*}{00:18:04}                                        & \multirow{2}{*}{03:07:19}     & \multirow{2}{*}{00:11:13}                           \\  
                                 &&&&&               
                     \\ 
                     \specialrule{.15em}{.15em}{.15em}
\end{tabular}
}}
\vspace{-0.15in}
\end{table}
\subsubsection{{\textbf{Running Time Comparison (RQ5)}}}
{Since we only model limited items for targets users in the re-ranking process (50 items for each user), it could be more efficient to model text information in the re-ranking process rather than modeling text information by a global ranking model as most review-aware ranking models do.     
However, modeling text reviews may result in significantly higher costs for RAISE compared to approaches that do not take reviews into account.
To investigate the computational costs in practical situations, we hence compare the running time of all methods as shown in Table \ref{tab:running time}. }

{We can observe that DLCM takes the most running time.
Meanwhile, RAISE takes much more time than PRM and SetRank due to the review modeling process happened in the IDM. 
Although modeling text review takes unexpected computational costs, it benefits the re-ranking model in threefold. 
\begin{itemize}[leftmargin=*]
\item First, IDM distinguishes the importance of reviews according to the intentions behind them, enabling RAISE to capture user-specific inter-item relationships and hence perform user-specific predictions. 
\item Second, modeling text information brings additional performance gains (see section 5.2.2). 
\item Third, RAISE is able to provide meaningful explanations for target users by modeling text information with the co-attention mechanism (see section 5.2.3).
\end{itemize}
}

{In summary, with more time spent, RAISE provides better personalization and interpretability than comparing methods by modeling text information.
In order to alleviate the above problem and decrease the training time, future works will introduce efficient attention mechanisms for the computations of co-attention and dynamic self-attention. For example, localizing the attention span and using memory-compressed attention are simple yet effective methods to decrease computational costs \cite{tay2020efficient}. }

\subsubsection{\textbf{{Effects of Text Review (RQ6)}}}

{To investigate the effect of user reviews and item reviews, we compare RAISE with its two variants listed as follows:
(1) RAISE$_\text{w/o u\_rvw}$: removing the user review information from RAISE;
(2) RAISE$_\text{w/o i\_rvw}$: removing the item review information from RAISE.}
{As shown in the Table \ref{tab:user review}, we can see the default setting of RAISE (i.e., considering both user and item reviews) performs the best on three out of four presented datasets. 
Surprisingly, the model performance on the video games dataset indicates that modeling item reviews only (i.e., RAISE$_\text{w/o u\_rvw}$) is able to achieve better results than RAISE which models both user and item reviews. This indicates the importance of modeling item reviews for the Video Games dataset.}

{Furthermore, we feed both text information and latent representation learned from implicit feedback into all re-ranking methods, to further investigate how much the text information can improve the model performance. 
Specifically, given an embedding review sequence $\mathcal{R}_{i}^e\!\!=\!\! \{\mathbf{r}_{1}^{(i)}, \mathbf{r}_{2}^{(i)}, \cdots, \mathbf{r}_{\textit{l}_i}^{(i)} \}$, we first obtain the representations of text information for each item by summing up its corresponding representation of reviews:}
\begin{equation}
\footnotesize
{ \mathbf{r}_{i} = \sum\nolimits_{j}  \mathbf{r}_{j}^{(i)}}
\end{equation}
{Then we concatenate the representations of text information $\mathbf{r}_{i}$ and item latent representation $\mathbf{q}_i$ to replace the original item representations $\mathbf{q}_i$ in the baseline re-ranking models. 
In addition, we implement the IDM for the PRM model to further evaluate the effectiveness of IDM since the PRM model achieves the best results on four datasets among all baseline methods, which is denoted as PRM$_{\text{IDM}}$.}
From Table \ref{tab:review_for_baselines}, we can see the model performance of review-aware baselines do not improve their performance as expected. 
In particular, considering text information significantly degrades the performance of DLCM on all datasets. 
This showcases that the model architectures of baselines should be modified accordingly in order to leverage reviews, and simply introducing textual information as input can lead to worse results.
Besides, the IDM-enhanced PRM (i.e., PRM$_{\text{IDM}}$) consistently outperforms PRM$_{review}$. 
Compared with PRM$_{review}$, it generates more meaningful review representations for the prediction model by computing matching scores between every user-item review pair with a co-attention network.
This helps confirm the effectiveness of the IDM module.

\begin{table}[!t]
\centering
\scriptsize
\caption{{The effects of text reviews of RAISE (\%).}}\label{tab:user review}
\vspace{-0.1in}
{
\setlength{\tabcolsep}{4pt}
{
\begin{tabular}{|c|c|cc|cc|cc|}
\specialrule{.15em}{.15em}{.15em}
\multirow{2}{*}{Dataset} &\multirow{2}{*}{Model} &\multicolumn{2}{c|}{Pre} &\multicolumn{2}{c|}{MAP} &\multicolumn{2}{c|}{NDCG} \\\cline{3-8}
& &@5 &@10 &@5 &@10 &@5 &@10 \\ \hline
                                                            \multirow{-1}{*}{\begin{tabular}[c]{@{}c@{}}Sports \\and \\Outdoors\end{tabular}}               & RAISE                                                       & \textbf{60.63}                               & \textbf{37.77}& \textbf{86.96}             & \textbf{83.48}                                       & \textbf{77.84}     & \textbf{80.30}                      \\ 
                                                                           & RAISE$_\text{w/o u\_rvw}$                                                      & 59.94                                             & 37.36                     & 86.30                                  & 82.85  & 77.00 	&79.10
\\ & RAISE$_\text{w/o i\_rvw}$                                                      & 59.04                                             & 37.25                     & 85.69                                  & 81.71  & 76.02 	&78.56
\\ \toprule
                                                            \multirow{-1}{*}{\begin{tabular}[c]{@{}c@{}}Health and\\Personal\\ Care\end{tabular}}               & RAISE                                                       & \textbf{59.48}                                                 & \textbf{36.73}   & \textbf{85.53}                                        & \textbf{82.21}         & \textbf{76.42}      & \textbf{78.99}                       \\  
                                                                           & RAISE$_\text{w/o u\_rvw}$                                                      & 57.86                                                 & 36.10                       &83.84                            & 80.85  & 75.03       	&77.77

\\ & RAISE$_\text{w/o i\_rvw}$                                                      & 57.87                                             & 36.13                     & 84.49                                  & 80.96  & 75.17 	&77.83                     
\\ \toprule
                                                            \multirow{-1}{*}{\begin{tabular}[c]{@{}c@{}}Clothing, \\Shoes and\\ Jewelry\end{tabular}}               & RAISE                                                       & \textbf{70.39}                                                 & \textbf{41.66}    & \textbf{95.32}                                        & \textbf{93.25}         & \textbf{91.33}      & \textbf{92.38}                       \\  
                                                                           & RAISE$_\text{w/o u\_rvw}$                                                      & 69.13                                                 & 41.07                       &94.79                            & 92.50  & 89.98       	&91.44

\\ & RAISE$_\text{w/o i\_rvw}$                                                      & 69.24                                             & 41.07                     & 94.69                                  & 92.40  & 90.01 	&91.39                     
                     
\\ \toprule
                                                                \multirow{-1}{*}{\begin{tabular}[c]{@{}c@{}}Video \\Games \end{tabular}}              & RAISE                                                       & 59.43                                                 & \textbf{40.68}    & 83.99                                        & 79.33         & 73.16      & 76.30                       \\  
                                                                           & RAISE$_\text{w/o u\_rvw}$                                                      & \textbf{59.81}                                                 & 40.51                       &\textbf{84.09}                            & \textbf{79.91}  & \textbf{73.61}       	&\textbf{76.55}
\\ & RAISE$_\text{w/o i\_rvw}$                                                      & 58.39                                             & 40.17                     & 82.58                                  & 78.09  & 71.85 	&75.24
                     \\ 
                     \specialrule{.15em}{.15em}{.15em}
\end{tabular}
}}
\vspace{-0.1in}
\end{table}

\begin{table}[!ht]
\centering
\scriptsize
\caption{{The performance comparison between RAISE and baselines with review information as input (\%).}}\label{tab:review_for_baselines}
\vspace{-0.1in}
{
\setlength{\tabcolsep}{5pt}
{
\begin{tabular}{|c|c|cc|cc|cc|}
\specialrule{.15em}{.15em}{.15em}
\multirow{2}{*}{Dataset} &\multirow{2}{*}{Model} &\multicolumn{2}{c|}{Pre} &\multicolumn{2}{c|}{MAP} &\multicolumn{2}{c|}{NDCG} \\\cline{3-8}
& &@5 &@10 &@5 &@10 &@5 &@10 \\ \hline

 & DLCM$_\text{review} $                                                   & 4.80                                             & 4.86                     & 9.09                                  & 9.71  & 5.23 	&6.57

        \\

                                                                           & SetRank$_\text{review} $                                                   & 49.48                                             & 33.89                     & 76.39                                  & 71.04  & 63.49 	&68.95
           \\                                                           
           
             & PRM$_\text{review} $                                                    & 50.60                                             & 34.31                    &  76.15                                 & 71.17  & 64.28 	&69.53
                                    \\                                       
                             & PRM$_\text{IDM} $                                                     &55.08                                             & 36.15                    &   82.55                                &77.71   &71.20  	&75.17                                                                          \\ 
                            \multirow{-4}{*}{\begin{tabular}[c]{@{}c@{}}Sports \\and\\ Outdoors\end{tabular}}                                                 & RAISE                                                       & \textbf{60.63}                               & \textbf{37.77}& \textbf{86.96}             & \textbf{83.48}                                       & \textbf{77.84}     & \textbf{80.30}

\\ \toprule
                 
                                                         \multirow{-1}{*}{\begin{tabular}[c]{@{}c@{}}Health \\and\\Personal\\ Care\end{tabular}}                     
                                                         & DLCM$_\text{review} $                                                     & 5.58                                             & 5.39                     & 11.82                                  & 12.18  & 6.43 	&7.82

        \\ 
                                                                       
                                                                           & SetRank$_\text{review}$                                                     & 46.68                                                 & 32.07                       &71.56                            & 67.41  & 59.63       	&65.50 
                                                                           \\
                                                                           & PRM$_\text{review} $                                                     &  49.38                                            & 33.53                     &   75.34                                & 70.65   & 63.28 	&68.80   \\                              & PRM$_\text{IDM} $                                                     &  53.76                                            & 35.03                     &   80.19                                & 76.07   & 69.85 	&73.97   \\

                                                                           & RAISE                                                       & \textbf{59.48}                                                 & \textbf{36.73}   & \textbf{85.53}                                        & \textbf{82.21}         & \textbf{76.42}      & \textbf{78.99}

\\ \toprule
                                                                                      & DLCM$_\text{review}$                                                     & 4.54                                                 & 4.21                       &8.76                            & 9.10  & 5.38       	&6.47
                                                                                      
        \\ 
                                                                           & SetRank$_\text{review} $                                                     & 58.22                                             & 37.91                    &   93.98                                & 79.66  & 76.16 	&81.21
                        \\
                                                                    
                                                            & PRM$_\text{review} $                                                     & 61.64                                             & 39.16                     & 86.72                                  & 83.01  & 80.33 	&84.63

                        \\  
                                     & PRM$_\text{IDM} $                                                     &66.41                                             & 40.60                    &   92.17                                &89.26   &86.69  	&89.30                  \\                            
                                                                                   & RAISE                                                       & \textbf{70.39}                                                 & \textbf{41.66}    & \textbf{95.32}                                        & \textbf{93.25}       & \textbf{91.33}      & \textbf{92.38}

\\ \toprule
                                                                      
                                                                 \multirow{0}{*}{\begin{tabular}[c]{@{}c@{}}Video \\Games \end{tabular}}             & DLCM$_\text{review}$                                                     & 20.25                                             & 19.84 & 32.50                                  & 32.91  & 21.91 	&28.09

        \\ 
                                  \multirow{-9}{*}{\begin{tabular}[c]{@{}c@{}}Clothing \\shoes\\and\\Jewelry\end{tabular}}                        
                                                                           & SetRank$_\text{review}$                                                     & 53.42                                                 & 38.15                       &77.02                            & 72.15  & 65.08       	&69.50 \\                      & PRM$_\text{review} $                                                     &51.79                                             & 37.51                    &   74.35                                &69.90   &62.52  	&67.52                  \\   
                                                                                        & PRM$_\text{IDM} $                                                     &55.85                                             & 39.52                    &   79.75                                &74.76   &68.22  	&72.61                  \\   
 & RAISE                                                       & \textbf{59.43}                                                 & \textbf{40.68}    & \textbf{83.99}                                        & \textbf{79.33}         & \textbf{73.16}      & \textbf{76.30}

                     \\ 
                     \specialrule{.15em}{.15em}{.15em}
\end{tabular}
}}
\vspace{-0.15in}
\end{table}

{One may wonder whether the number of reviews affects the model complexity. 
In our work, we set the number of reviews $\textit{l}_u$ and $\textit{l}_i$ to a constant (i.e., 20), and the text reviews are pre-processed before training.
In this way, RAISE loads user and item text representations for the training process, and the model complexity only grows with the number of users and items, and without regard to the number of reviews. 
In summary, additional memory costs are needed for training on larger datasets because RAISE needs to train and save the representations of users and items, which is very common for modern deep learning recommendation models.}
\subsubsection{\textbf{{Ablation Analysis for IDM (RQ7)}}}
{In the IDM, we adopt a bilinear co-attention function (Eq.~\ref{equ:co-attention}) to compute the matching scores, which enables RAISE to capture intention-aware information and to provide meaningful explanations.
We have tried two more co-attention function to compute the review-level matching scores in this section. 
The first one omits the transform matrix $\mathbf{M}$ and the latter one utilizes only one MLP to compute matching scores \cite{tay2018multi}.}
\begin{equation}
\footnotesize
{ c_{kj} = f(\mathbf{r}_{k}^{(u)})^{T} f(\mathbf{r}_{j}^{(i)})}
\end{equation}
\begin{equation}
\footnotesize
 {c_{kj} = f(Concat(\mathbf{r}_{k}^{(u)}, \mathbf{r}_{j}^{(i)}))}
\end{equation}
{The above co-attention functions are denoted as Co-ATT (Soft) and Co-ATT(MLP), respectively. 
We present the performance of Co-ATT (Soft) and Co-ATT (MLP) in the Table \ref{tab:equ5}. 
In addition, we change the aggregation function from the sum pooling to the mean pooling in order to discover the effect of different aggregation functions in the Eq.~\ref{equ:intention-representation}, which is denoted as "Aggr (Mean)" in the Table \ref{tab:equ5}. 
We observe the relative ranking of all three variants (bilinear, soft and MLP) are always interchanging across different datasets. 
On the other hand, changing the aggregation function could marginally improve performance on the later two datasets.}

\begin{table}[!h]
\centering
\scriptsize
\caption{{Abaltion analysis for IDM (\%). "Co-ATT" denotes the co-attention function used in the Equ. 3, and "Aggr" is short for aggregation function in the Equ. 5.}}\label{tab:equ5}
\vspace{-0.1in}
{
\setlength{\tabcolsep}{5pt}
{
\begin{tabular}{|c|c|cc|cc|cc|}
\specialrule{.15em}{.15em}{.15em}
\multirow{2}{*}{Dataset} &\multirow{2}{*}{Model} &\multicolumn{2}{c|}{Pre} &\multicolumn{2}{c|}{MAP} &\multicolumn{2}{c|}{NDCG} \\\cline{3-8}
& &@5 &@10 &@5 &@10 &@5 &@10 \\ \hline
                                                                         & RAISE                                                       & \textbf{60.63}                               & \textbf{37.77}& \textbf{86.96}             & \textbf{83.48}                                       & \textbf{77.84}     & \textbf{80.30}                  

        \\ 
                                                                           & Co-ATT (Soft)                                                     & 59.51                                             & 37.09                     & 85.85                                  & 82.28  & 76.44 	&78.56

        \\ 
                                                                           & Co-ATT (MLP)                                                     & 58.69                                             & 37.01                    &  85.26                                 & 81.39  & 75.52 	&78.07
                                                                           
                                                                               \\ \multirow{-4}{*}{\begin{tabular}[c]{@{}c@{}}Sports \\and \\Outdoors\end{tabular}}  
                                                                           & Aggr (Mean)                                                     & 59.57                                             & 37.37                     & 86.15                                  & 82.36  & 76.57 	&78.88
\\ \toprule
                                                                  & RAISE                                                       & \textbf{59.48}                                                 & 36.73   & \textbf{85.53}                                        & 82.21         & 76.42      & \textbf{78.99}                     

                   \\ 
                                                         \multirow{-2}{*}{\begin{tabular}[c]{@{}c@{}}Health\\ and\\Personal \\Care\end{tabular}}                           & Co-ATT (Soft)                                                     & 59.00                                             & \textbf{36.76}                     & 85.34                                  & \textbf{82.27}  & \textbf{76.46} 	&78.97

        \\ 
                                                                           & Co-ATT (MLP)                                                     &  58.10                                            &36.38                     &   84.39                                &81.26   & 75.40 	&78.15   \\  
                                                                           & Aggr (Mean)                                                     & 58.55                                                 & 36.40                       &84.76                            & 81.50  & 75.83       	&78.32       
\\ \toprule
                                                                      & RAISE                                                       & 70.39                                                 & 41.66    & 95.32                                        & 93.25       & 91.33      & 92.38                    

      \\ 
                                                            \multirow{-2}{*}{\begin{tabular}[c]{@{}c@{}}Clothing, \\ Shoes\\and\\ Jewelry\end{tabular}}                    & Co-ATT (Soft)                                                     & 70.18                                             & \textbf{41.77}                     & 95.58                                  & 93.21  & 91.06 	&92.40

        \\ 
                                                                           & Co-ATT (MLP)                                                     & 70.83                                             & 41.74                    &   \textbf{95.89}                                & 93.68  & 91.71 	&92.61
                      
                        \\  
                                                                           & Aggr (Mean)                                                     & \textbf{70.86}                                                 & 41.73                       &95.86                            & \textbf{93.74}  & \textbf{91.71}       	&\textbf{92.64}
\\ \toprule
                                                                           & RAISE                                                       & 59.43                                                 & \textbf{40.68}    & 83.99                                        & 79.33         & 73.16      & 76.30

                                                     \\ 
                                                                 \multirow{-1}{*}{\begin{tabular}[c]{@{}c@{}}Video \\Games \end{tabular}}             & Co-ATT (Soft)                                                     & 58.67                                             & 40.14                     & 82.99                                  & 78.51  & 72.17 	&75.44

        \\ 
                                                                           & Co-ATT (MLP)                                                     &\textbf{59.84}                                              & 40.46                    &   83.90                                &79.72   &\textbf{73.52}  	&\textbf{76.34}                  \\  
                                                                           & Aggr (Mean)                                                     & 59.72                                                 & 40.43                       &\textbf{84.23}                            & \textbf{79.75}  & \textbf{73.52}       	&76.32                  
                     \\ 
                     \specialrule{.15em}{.15em}{.15em}
\end{tabular}
}}
\vspace{-0.15in}
\end{table}
{
In summary, the default setting can help achieve the best performance on some datasets. 
For the other datasets, although the default setting do not perform the best, there is not huge performance difference between different settings.
As such, there is no universal settings for all datasets, and it may vary case by case.}

\section{Conclusion and Future Works}
In this paper, we propose a novel re-ranking method RAISE to refine the recommendation list. 
Equipped with the intention discover module (IDM) and dynamic transformer encoder (DTE), our proposed RAISE performs user-specific re-ranking by exploiting user intentions with the help of text reviews.
By constructing RAISE upon prior global ranking models, one can easily achieve personalization, efficiency, and interpretability without modifying their current recommendation engines. 
{Such scalability enables RAISE to refine recommendation lists generated by existing ranking models in an efficient manner.}
Empirical study verifies the additional gains brought by the devised IDM and DTE. 
In future work, we will investigate how to mine user intentions from other auxiliary information such as social networks and knowledge graphs for further performance-enhanced re-ranking approaches.
\section*{Acknowledgment}
This work was supported by Singapore Ministry of Education Academic Research Fund Tier 1 [2020-T1-001-130(RG15/20)], Singapore Ministry of Education Academic Research Fund Tier 2 [MOE2019-T2-2-175], and Singapore Institute of Manufacturing Technology-Nanyang Technological University (SIMTech-NTU) Joint Laboratory and Collaborative Research Programme on Complex Systems. 



\bibliographystyle{IEEEtran}
\bibliography{ref}

\begin{IEEEbiography}[{\includegraphics[width=1in,height=1.25in,clip,keepaspectratio]{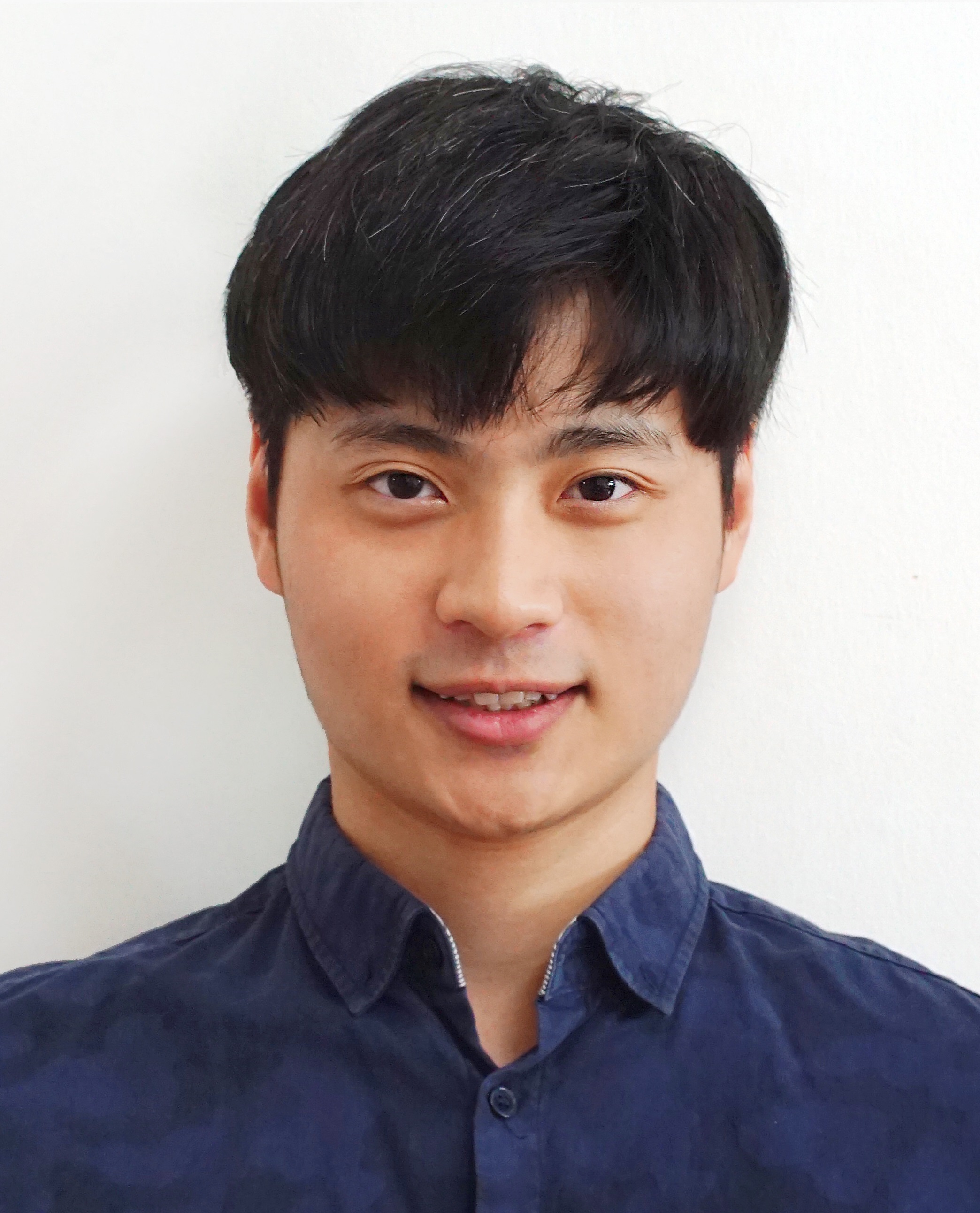}}]{Zhuoyi Lin}
is a research scientist in the Institute for Infocomm Research $(I^{2}R)$ at the Agency for Science, Technology and Research (A*STAR), Singapore.  
He is currently pursuing the Ph.D. degree with the School of Computer Science and Engineering, Nanyang Technological University, Singapore. He received the B.S. degree in Electronic Engineering from Ming Chuan University, Taiwan, in 2017. His research interests include collaborative filtering, deep learning and their applications.
\end{IEEEbiography}
\begin{IEEEbiography}[{\includegraphics[width=1in,height=1.25in,clip,keepaspectratio]{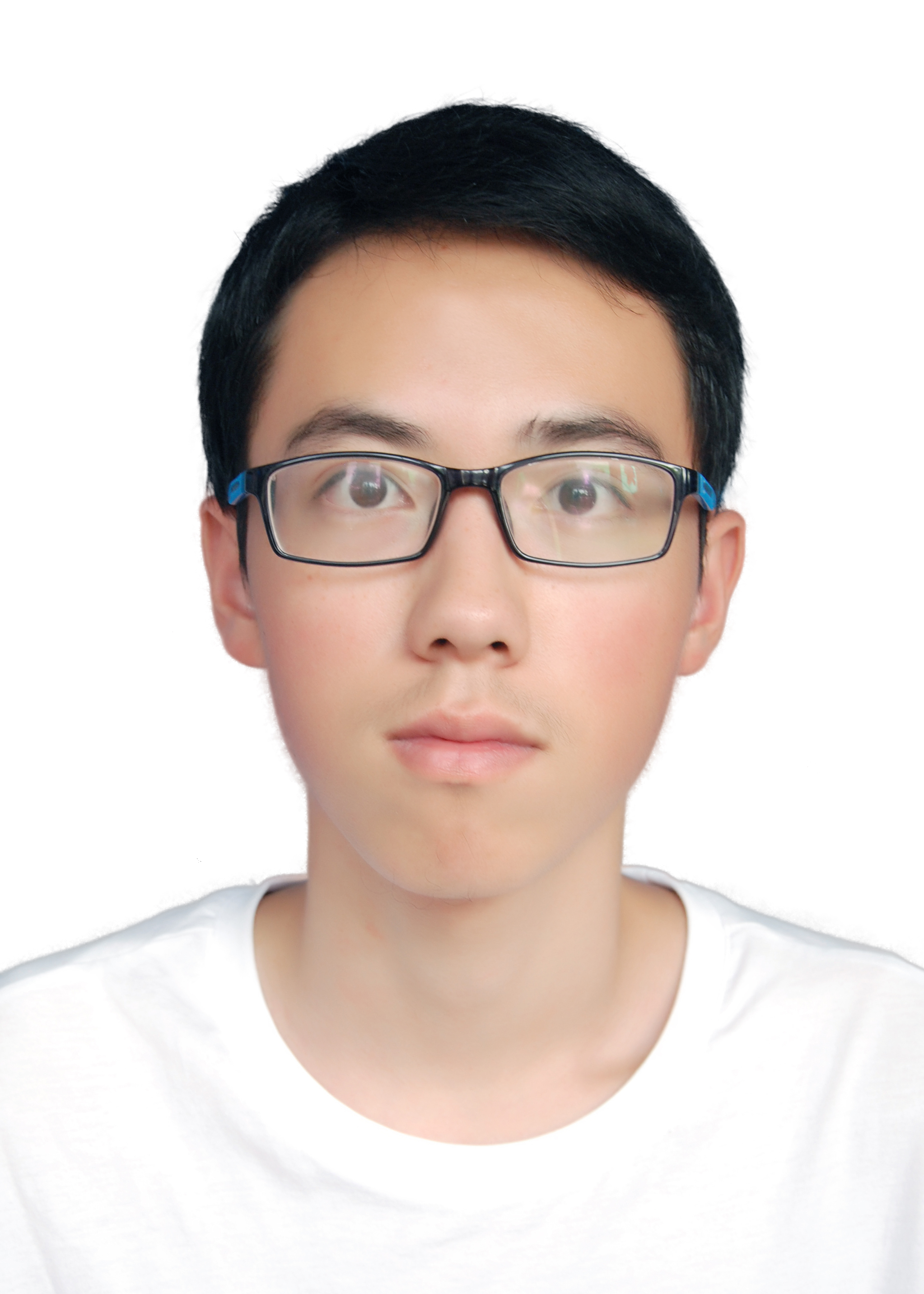}}]{Sheng Zang}
achieved his B.S. degree in Computer Science and Technology from Nankai University, China. His research interests mainly include deep learning, reinforcement learning and their applications. 
\end{IEEEbiography}
\begin{IEEEbiography}[{\includegraphics[width=1in,height=1.25in,clip,keepaspectratio]{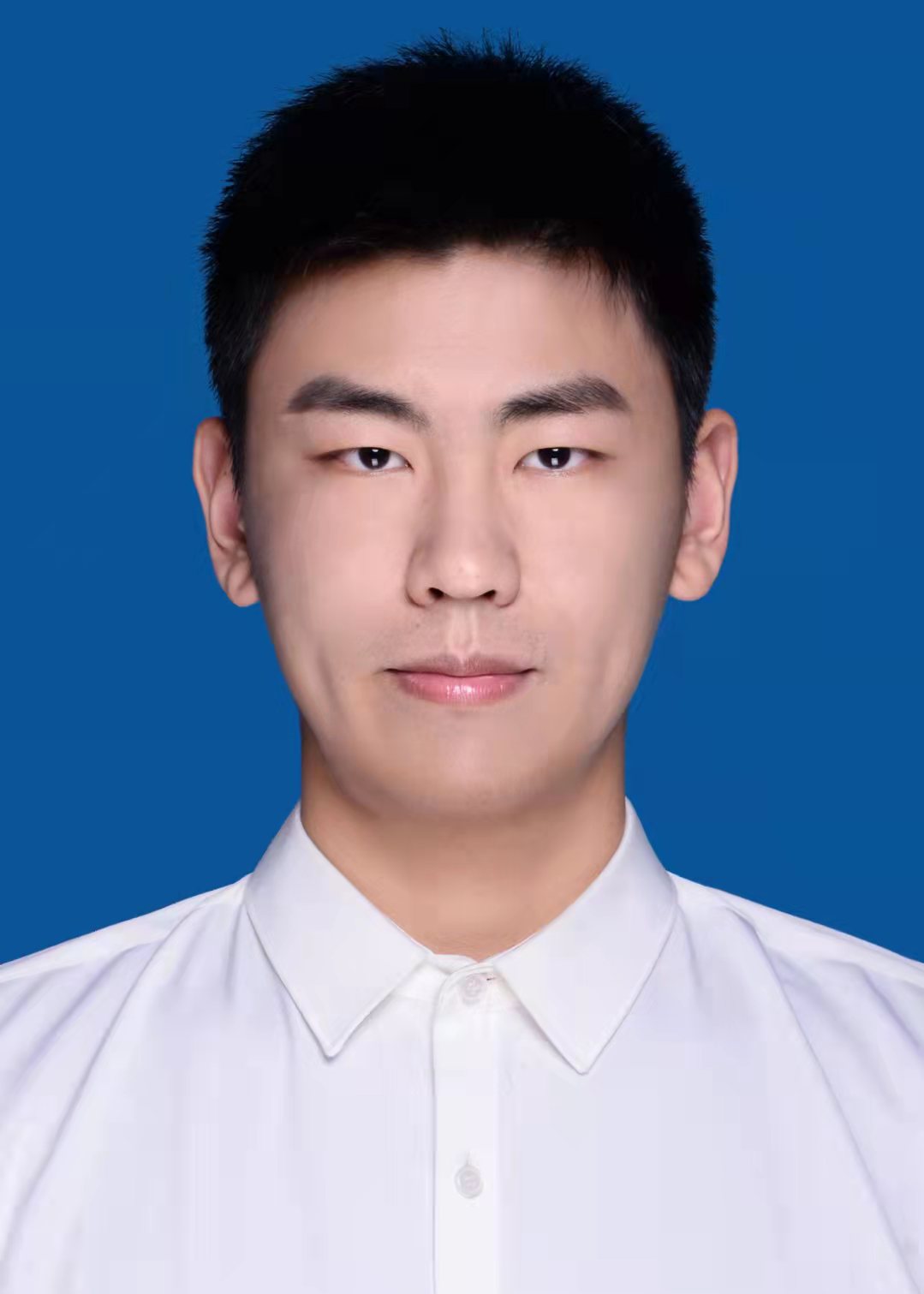}}]{Rundong Wang}
is a Ph.D. student at the School of Computer Science and Engineering, Nanyang Technological University. His research interest are (Multi-agent) Reinforcement Learning and application in Game, Finance, and so on. He has published several papers in the top AI conference such as ICML, ICLR, AAAI, IJCAI, etc.
\end{IEEEbiography}
\begin{IEEEbiography}[{\includegraphics[width=1in,height=1.25in,clip,keepaspectratio]{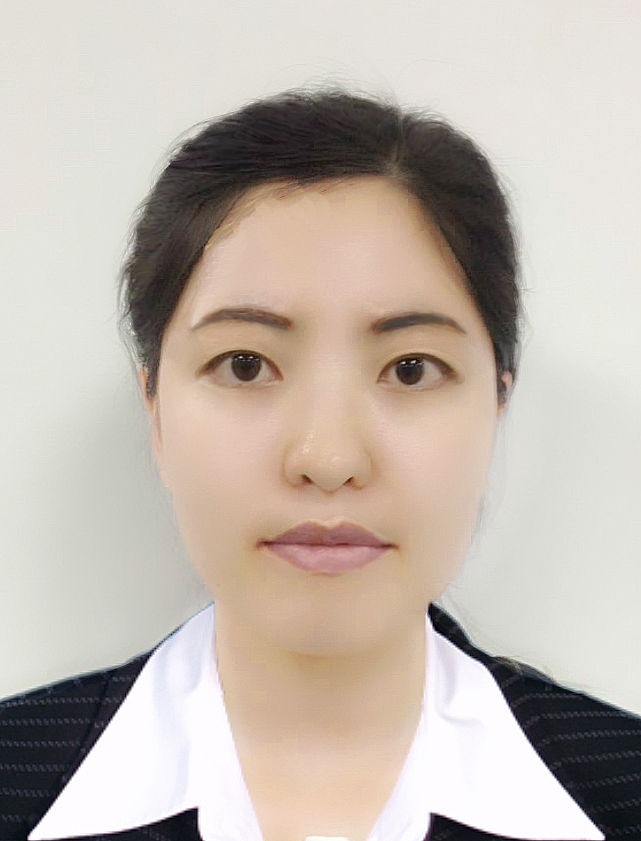}}]{Zhu Sun}
received her Ph.D. degree from Nanyang
Technological University, Singapore, in 2018. Her research interests includes machine learning, data mining and recommender systems. Her research has been published in leading conferences and journals in related domains (e.g., IJCAI, AAAI, CIKM, ACM RecSys, IEEE TKDE and TNNLS).
Currently, she is with Institute of High Performance Computing and Centre for
Frontier AI Research, A*STAR, Singapore. 
\end{IEEEbiography}
\begin{IEEEbiography}[{\includegraphics[width=1in,height=1.25in,clip,keepaspectratio]{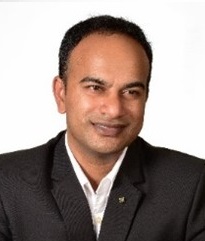}}]{J. Senthilnath} (Senior Member, IEEE) is a Scientist in the Institute for Infocomm Research $(I^{2}R)$ at the Agency for Science, Technology and Research (A*STAR), Singapore. He received the PhD degree in Aerospace Engineering from the Indian Institute of Science (IISc), India. His current research interests include artificial intelligence, multi-agent systems, generative models, online learning, reinforcement learning and optimization. He has published over 100 high-quality papers and won five best paper awards. He is a member of Artificial Intelligence, Analytics And Informatics (AI3), A*STAR. He has been serving as Guest Editor/organizing chair/co-chair/PC members in leading AI and data analytics journals and conferences.
\end{IEEEbiography}
\begin{IEEEbiography}[{\includegraphics[width=1in,height=1.25in,clip,keepaspectratio]{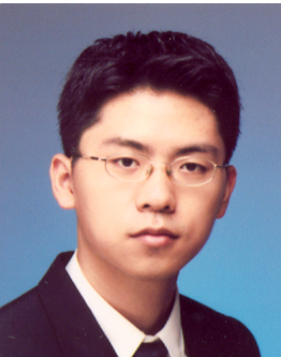}}]{Chi Xu}
received his Ph.D. and Bachelor (Honors) from Nanyang Technological University (NTU), Singapore, in 2010 and 2003 respectively. After joined SIMTech A*Star in 2010, he has been working in the areas of information management and operations analytics for enterprise and supply chain applications. He has developed various technologies which have been licensed to companies in different industries such as aerospace, construction, transportation\&logistics, and fast moving consumer goods (FMCG). Currently, he is deputy group manager of planning and operations management group (POM) of SIMTech. He is responsible to drive the group research direction and capabilities development for operation analytics and system optimization. 
He is an adjunct assistant professor of School of Computer Science and Engineering (SCSE) of Nanyang Technological University (NTU).
\end{IEEEbiography}
\begin{IEEEbiography}[{\includegraphics[width=1in,height=1.25in,clip,keepaspectratio]{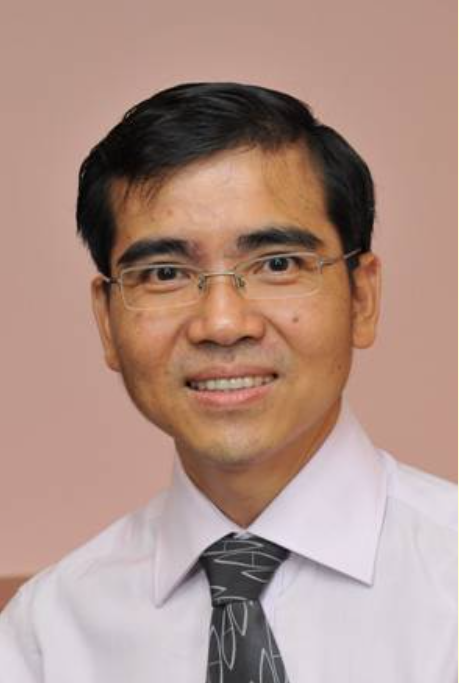}}]{Chee Keong Kwoh}
(Senior Member, IEEE) received the bachelor’s degree (Hons.) in electrical engineering and the master’s degree in industrial system engineering from the National University of Singapore, Singapore, in 1987 and 1991, respectively, and the Ph.D. degree from the Imperial College of Science, Technology and Medicine, University of London, in 1995.
He has been with the School of Computer Engineering, Nanyang Technological University (NTU), since 1993. 
His research interests include data mining, soft computing and graph-based inference; applications areas include bioinformatics and biomedical engineering. He has done significant research work in his research areas and has published many quality international conferences and journal articles. He has often been invited as an organizing member or referee and a reviewer for a number of premier conferences and journals, including GIW, IEEE, BIBM, RECOMB, PRIB, BIBE, ICDM, and iCBBE. He is also a member of the Association for Medical and Bioinformatics, Imperial College Alumni Association of Singapore. He has provided many services to professional bodies in Singapore and was conferred the Public Service Medal by the president of Singapore, in 2008. His research interests include data mining, soft computing and graph-based inference; applications areas include bioinformatics and biomedical engineering. He is an editorial board member of the International Journal of Data Mining and Bioinformatics, the Scientific World Journal, Network Modeling and Analysis in Health Informatics and Bioinformatics, Theoretical Biology Insights, and Bioinformation. He has been a Guest Editor of many journals, such as the Journal of Mechanics in Medicine and Biology, the International Journal on Biomedical and Pharmaceutical Engineering, and others.
\end{IEEEbiography}






\end{document}